\documentclass[pra,aps,showpacs,onecolumn,superscriptaddress]{revtex4-1}

\usepackage{amsmath,amsfonts,amssymb,color,epsfig,graphics,graphicx,latexsym,revsymb,theorem,url,verbatim}

\usepackage{hyperref}

\newtheorem{definition}{Definition}
\newtheorem{proposition}[definition]{Proposition}
\newtheorem{lemma}[definition]{Lemma}

\newtheorem{theorem}[definition]{Theorem}
\newtheorem{corollary}[definition]{Corollary}
\newtheorem{conjecture}[definition]{Conjecture}

\newtheorem{remark}[definition]{Remark}
\newtheorem{example}[definition]{Example}
\newtheorem{question}[definition]{Question}

\def\squareforqed{\hbox{\rlap{$\sqcap$}$\sqcup$}}
\def\qed{\ifmmode\squareforqed\else{\unskip\nobreak\hfil
\penalty50\hskip1em\null\nobreak\hfil\squareforqed
\parfillskip=0pt\finalhyphendemerits=0\endgraf}\fi}
\def\endenv{\ifmmode\;\else{\unskip\nobreak\hfil
\penalty50\hskip1em\null\nobreak\hfil\;
\parfillskip=0pt\finalhyphendemerits=0\endgraf}\fi}
\newenvironment{proof}{\noindent \textbf{{Proof.~} }}{\qed}
\def\Dbar{\leavevmode\lower.6ex\hbox to 0pt
{\hskip-.23ex\accent"16\hss}D}
\makeatletter
\def\url@leostyle{%
  \@ifundefined{selectfont}{\def\UrlFont{\sf}}{\def\UrlFont{\small\ttfamily}}}
\makeatother
\def\bcj{\begin{conjecture}}
\def\ecj{\end{conjecture}}
\def\bcr{\begin{corollary}}
\def\ecr{\end{corollary}}
\def\bd{\begin{definition}}
\def\ed{\end{definition}}
\def\bea{\begin{eqnarray}}
\def\eea{\end{eqnarray}}
\def\bem{\begin{enumerate}}
\def\eem{\end{enumerate}}
\def\bex{\begin{example}}
\def\eex{\end{example}}
\def\bim{\begin{itemize}}
\def\eim{\end{itemize}}
\def\bl{\begin{lemma}}
\def\el{\end{lemma}}
\def\bpf{\begin{proof}}
\def\epf{\end{proof}}
\def\bpp{\begin{proposition}}
\def\epp{\end{proposition}}
\def\bqu{\begin{question}}
\def\equ{\end{question}}
\def\br{\begin{remark}}
\def\er{\end{remark}}
\def\bt{\begin{theorem}}
\def\et{\end{theorem}}

\def\btb{\begin{tabular}}
\def\etb{\end{tabular}}

\newcommand{\nc}{\newcommand}

\def\a{\alpha}
\def\b{\beta}
\def\g{\gamma}

\def\m{\mu}

\def\r{\rho}
\def\s{\sigma}

\def\ps{\psi}

\def\G{\Gamma}
\def\D{\Delta}
\def\T{\Theta}
\def\L{\Lambda}

 \nc{\bA}{{\bf A}} \nc{\bB}{{\bf B}} \nc{\bC}{{\bf C}}
 \nc{\bD}{{\bf D}} \nc{\bE}{{\bf E}} \nc{\bF}{{\bf F}}
 \nc{\bG}{{\bf G}} \nc{\bH}{{\bf H}} \nc{\bI}{{\bf I}}
 \nc{\bJ}{{\bf J}} \nc{\bK}{{\bf K}} \nc{\bL}{{\bf L}}
 \nc{\bM}{{\bf M}} \nc{\bN}{{\bf N}} \nc{\bO}{{\bf O}}
 \nc{\bP}{{\bf P}} \nc{\bQ}{{\bf Q}} \nc{\bR}{{\bf R}}
 \nc{\bS}{{\bf S}} \nc{\bT}{{\bf T}} \nc{\bU}{{\bf U}}
 \nc{\bV}{{\bf V}} \nc{\bW}{{\bf W}} \nc{\bX}{{\bf X}}
 \nc{\bZ}{{\bf Z}}

\nc{\cA}{{\cal A}} \nc{\cB}{{\cal B}} \nc{\cC}{{\cal C}}
\nc{\cD}{{\cal D}} \nc{\cE}{{\cal E}} \nc{\cF}{{\cal F}}
\nc{\cG}{{\cal G}} \nc{\cH}{{\cal H}} \nc{\cI}{{\cal I}}
\nc{\cJ}{{\cal J}} \nc{\cK}{{\cal K}} \nc{\cL}{{\cal L}}
\nc{\cM}{{\cal M}} \nc{\cN}{{\cal N}} \nc{\cO}{{\cal O}}
\nc{\cP}{{\cal P}} \nc{\cQ}{{\cal Q}} \nc{\cR}{{\cal R}}
\nc{\cS}{{\cal S}} \nc{\cT}{{\cal T}} \nc{\cU}{{\cal U}}
\nc{\cV}{{\cal V}} \nc{\cW}{{\cal W}} \nc{\cX}{{\cal X}}
\nc{\cZ}{{\cal Z}}

\nc{\hA}{{\hat{A}}} \nc{\hB}{{\hat{B}}} \nc{\hC}{{\hat{C}}}
\nc{\hD}{{\hat{D}}} \nc{\hE}{{\hat{E}}} \nc{\hF}{{\hat{F}}}
\nc{\hG}{{\hat{G}}} \nc{\hH}{{\hat{H}}} \nc{\hI}{{\hat{I}}}
\nc{\hJ}{{\hat{J}}} \nc{\hK}{{\hat{K}}} \nc{\hL}{{\hat{L}}}
\nc{\hM}{{\hat{M}}} \nc{\hN}{{\hat{N}}} \nc{\hO}{{\hat{O}}}
\nc{\hP}{{\hat{P}}} \nc{\hR}{{\hat{R}}} \nc{\hS}{{\hat{S}}}
\nc{\hT}{{\hat{T}}} \nc{\hU}{{\hat{U}}} \nc{\hV}{{\hat{V}}}
\nc{\hW}{{\hat{W}}} \nc{\hX}{{\hat{X}}} \nc{\hZ}{{\hat{Z}}}

\nc{\hn}{{\hat{n}}}

\def\aff{\mathop{\rm aff}}

\def\diag{\mathop{\rm diag}}
\def\dim{\mathop{\rm Dim}}

\def\ghz{\mathop{\rm GHZ}}

\def\lin{\mathop{\rm span}}

\def\rank{\mathop{\rm rank}}

\def\tr{\mathop{\rm Tr}}

\def\GL{{\mbox{\rm GL}}}

\def\SL{{\mbox{\rm SL}}}

\def\Un{{\mbox{\rm U}}}

\def\ox{\otimes}

\def\pars{\partial\cS}

\def\pard{\partial\cD}

\def\su{\subset}
\def\sue{\subseteq}

\def\we{\wedge}
\newcommand{\bra}[1]{\langle#1|}
\newcommand{\ket}[1]{|#1\rangle}
\newcommand{\proj}[1]{| #1\rangle\!\langle #1 |}
\newcommand{\ketbra}[2]{|#1\rangle\!\langle#2|}

\newcommand{\norm}[1]{\lVert#1\rVert}
\newcommand{\abs}[1]{|#1|}

\newcommand{\jmp}{J. Math. Phys.}

\def\Dbar{\leavevmode\lower.6ex\hbox to 0pt
{\hskip-.23ex\accent"16\hss}D}

\begin{document}
\title{Boundary of the set of separable states}

\author{Lin Chen}
\email{linchen@buaa.edu.cn (corresponding author)}
\affiliation{School of Mathematics and Systems Science, Beihang University, Beijing 100191, China}
\affiliation{International Research Institute for Multidisciplinary Science, Beihang University, Beijing 100191, China}

\def\Dbar{\leavevmode\lower.6ex\hbox to 0pt
{\hskip-.23ex\accent"16\hss}D}
\author {{ Dragomir {\v{Z} \Dbar}okovi{\'c}}}
\email{djokovic@uwaterloo.ca}
\affiliation{Department of Pure Mathematics and Institute for
Quantum Computing, University of Waterloo, Waterloo, Ontario, N2L
3G1, Canada} 

\begin{abstract}
Motivated by the separability problem in quantum systems 
$2\ox 4$, $3\ox 3$ and $2\ox 2\ox 2$, we study the maximal 
(proper) faces of the convex body, $\cS_1$, of normalized 
separable states in an arbitrary quantum system with 
finite-dimensional Hilbert space $\cH=\cH_1\ox\cH_2\ox\cdots\ox\cH_n$. To any subspace $V\sue\cH$ we associate a face $F_V$ of $\cS_1$ consisting of all states $\r\in\cS_1$ whose range is contained in $V$. We prove that $F_V$ is a maximal face if and only if $V$ is a hyperplane. If $V=\ket{\psi}^\perp$ where $\ket{\psi}$ is a product vector, we prove that $\dim F_V=d^2-1-\prod(2d_i-1)$, where $d_i=\dim \cH_i$ and $d=\prod d_i$. We classify the maximal faces of $\cS_1$ in the cases $2\ox 2$ and $2\ox 3$. In particular we show that the minimum and the maximum dimension of maximal faces is 6 and 8 for $2\ox2$, and 20 and 24 for $2\ox3$. The boundary, $\pars_1$, of $\cS_1$ is the union of all maximal faces. When $d>6$ it is easy to show that there exist 
full states on $\pars_1$, i.e., states $\r\in\pars_1$ such that all partial transposes of $\r$ (including $\r$ itself) have rank $d$. 
K.-C. Ha and S.-K. Kye have recently constructed explicit such states in $2\ox4$ and $3\ox3$. In the latter case, they have also 
constructed a remarkable family of faces, depending on a real parameter $b>0$, $b\ne1$. Each face in the family is a 9-dimensional simplex and any interior point of the face is a full state. We construct suitable optimal entanglement witnesses (OEW) 
for these faces and analyze the three limiting cases $b=0,1,\infty$.
\end{abstract}


\maketitle

Keyword: separable state, partial transpose, maximal face, entanglement witness

\tableofcontents

\section{Introduction}

Let $\cH=\cH_1\otimes\cH_2\otimes\cdots\otimes\cH_n$ be the
complex Hilbert space of a finite-dimensional $n$-partite quantum
system. We denote by $d_i$ the dimension of $\cH_i$, and so
$d:=\prod d_i$ is the dimension of $\cH$. To avoid trivial cases, we assume that each $d_i>1$ and $n>1$. Let $H$ be the space of
Hermitian operators $\r$ on $\cH$. Note that $H$ is a real vector space and that $\dim H=d^2$. We denote by $H_1$ the affine
hyperplane of $H$ defined by the equation $\tr\r=1$.
The mixed quantum states of this quantum system are represented
by their density matrices, i.e., operators $\r\in H$ which are
positive semidefinite $(\r\ge0)$ and have unit trace
$(\tr \r=1)$. For convenience, we often work with non-normalized
states, i.e., Hermitian operators $\r$ such that $\r\ge0$ and
$\r\ne0$. It will be clear from the context whether we require
the states to be normalized.

We denote by $\cD_1$ and $\cD$ the set of normalized and
non-normalized states, respectively. Thus $\cD_1=\cD\cap H_1$
is a compact convex subset of $H_1$.
We say that an operator $\r\in H$ {\em has full rank} if it is
invertible, and otherwise we say that $\r$ {\em has deficient
rank}. The boundary $\pard$ of $\cD$ (as a subset of $H$) consists of the zero operator and all states of deficient rank, i.e.,
 \bea
\pard=\{\r\in H:\r\ge0,~\rank \r<d\}.
 \eea

We say that a nonzero vector $\ket{a}\in\cH$ is a {\em product
vector} (or that it is {\em separable}) if it is the tensor
product $\ket{a}=\ket{a_1}\otimes\cdots\otimes\ket{a_n}$ of
vectors $\ket{a_i}\in\cH_i$. For brevity, we also write
it as $\ket{a}=\ket{a_1,\ldots,a_n}$. Otherwise we say that
$\ket{a}$ is {\em entangled}. A state $\r$ is a
{\em pure product state} if $\r=\proj{a}$ for some product vector
$\ket{a}$. A state $\s$ is {\em separable} if it can be written as a sum of pure product states. We shall denote by
$\cS_1$ and $\cS$ the set of normalized and non-normalized
separable states, respectively. Note that both $\cD\cup\{0\}$ and
$\cS\cup\{0\}$ are closed convex cones and $\cS\sue\cD$. We say
that a state is {\em entangled} if it is not separable. While
$\pard$ has a very simple description, the boundary $\pars$ of $\cS$ is well understood only for $d\le6$. That is, a two-qubit
or qubit-qutrit separable state belongs to $\pars$ if and only if
it or its partial transpose has deficient rank. The boundary
$\pars_1$ of $\cS_1$ (as a subset of $H_1$) and $\pars$ are
closely related. Indeed, we have
$\pars=\{t\r:\r\in\pars_1,t>0\}$.
The set $\cS_1$ is the convex hull of the set of all normalized
pure product states. Moreover, the latter set is the set of
extreme points of $\cS_1$.

For any vector subspace $V\sue\cH$, we denote by $P_V$ the set of
normalized product vectors contained in $V$. In particular,
$P_\cH$ is the set of all normalized product vectors in $\cH$.  Note that $P_V=V\cap P_\cH$ for any vector subspace $V\sue\cH$.
Finally, we set $R_V:=\{\proj{v}:\ket{v}\in P_V\}$. Thus,
$R_\cH$ is the set of all normalized pure product states. The range of a linear operator $\r$ will be denoted by $\cR(\r)$.

The partial transposition operators, $\G$, form an elementary 
Abelian group, $\T$, of order $2^n$. These operators act on the
algebra $\cA$ of all complex linear operators on $\cH$. Their
definition depends on the choice of bases in the Hilbert spaces
$\cH_i$. We assume that some orthonormal (o.n.) basis
$\{\ket{j}:0\le j<d_i\}$ of $\cH_i$ is fixed for each $i$. We use
the fact that $\cA$ is the tensor product of the algebras
$\cA_i$, $i=1,\ldots,n$, of linear operators on $\cH_i$. Thus
each $A\in\cA$ can be written as a finite sum of so called
{\em local operators} (LO), i.e., operators of the form
$\otimes_{i=1}^d A_i$, $A_i\in\cA_i$. The partial transposition
operator $\G_j$, $j=1,\ldots,n$, is characterized by the property
that it sends $\otimes_{i=1}^n A_i\to\otimes_{i=1}^n B_i$, where
$B_i=A_i$ for $i\ne j$ and $B_j=A_j^T$ is the transpose of $A_j$
(computed in our fixed o.n. basis of $\cH_j$). Obviously,
$\G_i\G_j=\G_j\G_i$ for all $i$ and $j$. We define $\T$ to be the
group generated by the $\G_i$s. For convenience, we set
$\r^\G=\G(\r)$. We say that a Hermitian operator $\r\in H$ is
{\em full} if  $\r^\G$ has full rank for all $\G\in\T$. 

We say that a state $\r$ on $\cH$ {\em has positive partial
transposes} (or that it is a {\em PPT state}) if $\r^\G\ge0$ for
all $\G\in\T$. We denote by $\cP$ the cone consisting of all
non-normalized PPT states, and we set $\cP_1=\cP\cap H_1$.
It follows that $\cP_1=\cap_{\G\in\T}\G(\cD_1)$.
Since each $\G\in\T$ preserves the set $R_\cH$ of normalized 
pure product states, it also preserves the set $\cS_1$. 
In general we have $\cS_1\sue\cP_1\su\cD_1$ and the equality 
$\cS_1=\cP_1$ holds if and only if $d\le6$.

We recall some basic notions and terminology concerning compact
convex subsets of a Euclidean space. In our case this space
will be $H$ or its affine subspace $H_1$. Occasionally we
shall apply this terminology to more general convex sets
such as $\cD,\cP$ and $\cS$.

Let $K\sue H_1$ be any compact convex subset. We denote by $\aff(K)$ the smallest affine subspace of $H_1$ containing $K$. By definition,
the dimension of $K$ is equal to the dimension of $\aff(K)$. The
relative boundary of $K$, i.e. as a subset of $\aff(K)$, will be
denoted by $\partial K$. A {\em face} of $K$ is a convex subset
$F\sue K$ such that the conditions $x,y\in K$ and $px+(1-p)y\in F$, $0<p<1$, imply that $x,y\in F$. The set $K$ itself is its own face, the unique {\em improper face}, all other faces are called {\em proper faces}. We say that a proper face $F$ of $K$ is {\em exposed} if there exists an affine hyperplane $
\su H_1$ such that $X\cap K=F$. Since $K$ is compact, it is clear that the empty face is exposed. By convention, the improper face is also 
exposed. A face $F$ of $K$ is {\em maximal} if there is no face $F'$ such that $F\su F'\su K$ and $F\ne K$. The boundary $\partial K$ is the union of all maximal faces of $K$. Given any subset $X\sue K$, there is the smallest (with respect to inclusion) face $F$ of $K$ containing $X$, and we say that the face $F$ is {\em generated} by $X$.

This paper is motivated by the desire to solve the separability
problem for some low-dimensional quantum systems such as
$2\otimes4$, $3\otimes3$, and $2\otimes2\otimes2$. In our
previous publication \cite[p. 5]{cd13JMP} we have proposed a
method to do that based on the theory of invariants. The recent
paper of P. D. Jarvis \cite{ja13} can be viewed as a first step
in that direction. To make further progress, it is necessary to obtain a better understanding of the boundary of the set $\cS_1$. Our objective is to present some basic facts (old and new) concerning $\pars_1$ and raise some related challenging problems.

Let $\Phi_k:H\to\bR$, $k=1,\ldots,d$, be the polynomial 
functions defined as follows: $\Phi_k(\r)$ is the sum of all 
$k\times k$ principal minors of the matrix $\r\in H$. Note 
that $\Phi_1=\tr$. We recall that the affine hyperplane $H_1$ is defined by the equation $\tr(\r)=1$. The convex body $\cD_1$ can be described as the set of all ponts $\r\in H_1$ satisfying $d-1$ inequalities
 \bea \label{eq:Nejednakosti-1}
\Phi_k(\r)\ge0,\quad k=2,\ldots,d.
 \eea

We can also define the convex body $\cP_1$ by a bunch of polynomial inequalities. For each $\G\in\T$ we define the polynomial function $\Phi_k^\G:H\to\bR$ by setting 
$\Phi_k^\G(\r)=\Phi_k(\r^\G)$. Then the set $\cP_1$ can be described as the set of all ponts $\r\in H_1$ satisfying 
$2^n(d-1)$ inequalities
 \bea \label{eq:Nejednakosti-2}
\Phi_k^\G(\r)\ge0,\quad \G\in\T,~ k=2,\ldots,d.
 \eea

The faces of $\cD_1$ are parametrized by vector subspaces
$V\sue\cH$ \cite[section II]{as10}. The face $\tilde{F}_V$ that corresponds to $V$ consists of all states $\r\in\cD_1$ such that $\cR(\r)\sue V$.
The intersection 
 \bea \label{eq:FaceF_V}
F_V:=\tilde{F}_V\cap\cS_1=\{\r\in\cS_1:\cR(\r)\sue V\}
 \eea
is a face (possibly empty) of $\cS_1$. We say that the face $F_V$ is {\em associated} to $V$. As each face of $\cD_1$ is exposed, it follows that each $F_V$ is an exposed face of $\cS_1$.

We denote by $\cF$ the set of all faces of $\cS_1$. Each $\G\in\T$ preserves $\cS_1$, permutes the faces of $\cS_1$ and preserve their properties. For any $F\in\cF$, we denote by $P_F$ the set of product vectors $\ket{z}\in\cH$ such that $\proj{z}\in F$. It is immediate from the definitions that for any subspace $V\sue\cH$ we have 
 \bea \label{eq:P_{F_V}} 
P_V=P_{F_V}.
 \eea

For any $F\in\cF$ we set $\cR(F)=\sum_{\r\in F}\cR(\r)$. Thus $\cR(F)$ is the smallest subspace of $\cH$ which contains $\cR(\r)$ for all $\r\in F$. It is easy to verify that $\cR(F)=\lin P_F$ for any $F\in\cF$. Note that there always exists $\r\in F$ such that $\cR(\r)=\cR(F)$. Further, we have
 \bea \label{eq:FandF'}
F\sue F_{\cR(F)}, \quad \forall F\in\cF.
 \eea
Indeed, for $\r\in F$ we have $\cR(\r)\sue\cR(F)$ and so 
$\r\in F_{\cR(F)}$. The inclusion in \eqref{eq:FandF'} may be proper (see  Example \ref{ex:FinF'}).

We say that $F\in\cF$ is an {\em induced face} if $F=\G(F_V)$ for some subspace $V\sue\cH$ and some $\G\in\T$. Since each face
$F_V$ is exposed, the same is true for all induced faces.
One defines the induced faces of $\cS$ similarly. We shall see later (Proposition \ref{pp:NonTrivial}) that when $d>6$ there exist maximal faces which are not induced. We warn the reader that our definition of induced faces is different from the one 
adopted in \cite{hk14} in the bipartite case. It is easy to see that in the bipartite case every induced face $F\in\cF$ is also induced according to the definition in that paper. However, the converse is false. A counter-example is provided by the face $F'$ of Proposition \ref{pp:RealVersion} when $n=2$ and $d_1=d_2=2$.

If $F\in\cF$ is a proper induced face then $F=\G(F_V)$ for some subspace $V\su\cH$ and some $\G\in\T$. Thus, $\G(F)=F_V\su\pard_1$ and at least one of the functions $\Phi_k^\G$, $k=2,\ldots,d$,  must vanish on $F$. It follows that the union of all proper 
induced faces of $\cS_1$ is equal to $\partial\cP_1\cap\cS_1$. The points $\r\in\cS_1$ satisfy all inequalities \eqref{eq:Nejednakosti-2} but if $d>6$ they must also satisfy some additional inequalities because $\cS_1\su\cP_1$. In order to find these additional inequalities we need to construct some non-induced faces $F$ of $\cS_1$, namely those which are not contained in $\partial\cP_1$. Note that if $\r\in F\in\cF$ and $\r\notin\partial\cP_1$ then $\r$ must be a full state. We shall consider such states in the next section.

Among the three smallest systems with $d>6$, namely $2\ox4$,
$3\ox3$ and $2\ox2\ox2$, proper faces $F\in\cF$ not contained 
in $\partial\cP_1$ are known only in $3\ox3$. A remarkable family of such faces, $\D_b$, depending on the real parameter $b>0$, 
$b\ne1$, has been constructed recently in \cite{hk14}, see 
also Example \ref{ex:FinF'} below. We shall study this family in section \ref{sec:3x3}. In the case $2\ox2\ox2$, no concrete full states on the boundary of $\cS_1$ are known.

In section \ref{sec:Bdry} we study the separable full states lying on the boundary of $\cS_1$. In Proposition \ref{pp:ParTransp} we show that if $\r\in\cS_1$ is not a full state, then $\r\in\pars_1$. It follows that such $\r$ belongs to a proper induced face of $\cS_1$. In Proposition 
\ref{pp:NonTrivial} we show that if a proper face $F$ of $\cS_1$ contains a full state, then $F$ is not induced. We deduce that $\cS_1\cap\partial\cP_1$ is the union of all proper induced faces of $\cS_1$. In the same proposition we also prove that if $d>6$ then there exist full states lying on $\pars_1$. 
In the $2\ox4$ and $3\ox3$ systems infinitely many explicit full states on $\pars_1$ are known (see \cite{hk2014} and Example \ref{ex:FinF'}). In the case $3\ox3$, each face $\D_b$ mentioned above is a 9-dimensional simplex and each interior point of 
$\D_b$ is a full state. However, in the case $2\ox4$ we can only 
say that the smallest face $F\in\cF$ containing one of the chosen 
full states is not contained in $\partial\cP_1$. For instance, we cannot determine the affine subspace spanned by $F$ and, in 
particular, we do not know the dimension of $F$. Ideally, one would like to know also the set of extreme points of $F$ (i.e., 
the set of pure product states contained in $F$).

In section \ref{sec:Some} we consider the special case where $d_1=\cdots=d_n$ and we use the identifications $\cH_1=\cdots=\cH_n$. 
We denote by $\cH_{\rm sym}$ the subspace of symmetric tensors 
of $\cH$. In Proposition \ref{pp:SymOper} we compute the dimension of the face $F=F_{\cH_{\rm sym}}$ and show that its extreme points are the states $\proj{x,\ldots,x}$ where $\ket{x}\in\cH_1$ and $\|x\|=1$. In Proposition \ref{pp:RealVersion} we determine the extreme points of the face $\cap_{\G\in\T}\G(F)$ as well as the subspace of $H$ spanned by this face. The dimension of this subspace is given by Eq. \eqref{eq:dimHsT}, and $\dim F$ is just one less.

In section \ref{sec:InducedMax} we introduce the $\diamond$-action of $\GL$ on $\cD_1$ which sends $\r\in\cD_1$ to the normalization of $A\r A^\dag$. One of the main results of this section is Proposition \ref{pp:maxfacedim} where we prove that for every maximal face $F$ of $\cS_1$ we have $\dim\cR(F)\ge d-1$. This implies that a maximal face $F$ is induced if and only if $\dim\cR(\G(F))=d-1$ for some $\G\in\T$. The second main result of this section (Theorem \ref{thm:hyperplane}) is that the face $F_V$, associated to a subspace $V\sue\cH$, is maximal if and only if $V$ is a hyperplane. As a corollary we obtain that every hyperplane 
$V\su\cH$ is spanned by product vectors. We compute the dimension of any maximal face $F_V$ where $V=\ket{\a}^\perp$ in two cases: first for arbitrary $n$ with $\ket{\a}$ a product vector and second for $n=2$ with $\ket{\a}$ of Schmidt rank two. We also obtain a very simple classification of maximal faces in $2\ox2$ and $2\ox3$ up to the $\diamond$-action of $\GL$.

In section \ref{sec:3x3} we study the above mentioned family $\D_b$, $b>0$, $b\ne1$, of 9-dimensional faces of $\cS_1$ in the $3\ox3$ system. We include the limiting cases $b=0,1,\infty$. We construct entanglement witnesses (EW) $W_b$ such that 
$\D_b=\{\r\in\cS_1:\tr(\r W_b)=0\}$. When $b\ne1$ then $W_b$ 
is in fact an OEW. For $b\ne1$, each interior point of $\D_b$ is a full state and so the face $\D_b$ is not induced. As a 
by-product of these results we obtain that the set of normalized 
(i.e., having trace 1) OEW is not closed. This family of states 
is remarkable as it provides the first examples of non-induced faces of relatively high dimension. No such faces are known 
in $2\ox4$ and $2\ox2\ox2$ systems. In the $2\ox4$ case, full 
states on the boundary of $\cS_1$ have been constructed very recently \cite{hk2014}.


\section{Boundary of the set of separable states}
\label{sec:Bdry}

\subsection{Basic facts}

It is well known that $\cS_1$ is a compact convex set and
that $\dim\cS_1=d^2-1$. Let us begin with three observations concerning the boundary of $\cS_1$.

{\em Observation 1}. $\cS_1$ is not a polytope.

In the bipartite case this been shown in \cite{it06}. While the
proof given there can be easily extended to the multipartite
case, we shall give a slightly different and shorter proof for
the general case. Clearly $\cS_1$ has at least one extreme point.
As $\cS_1$ is the convex hull of $R_\cH$, every extreme point of
$\cS_1$ must belong to $R_\cH$. Since the local unitary group
$\times_{i=1}^n \Un(d_i)$ acts transitively on $R_\cH$, each
$\r\in R_\cH$ is an extreme point of $\cS_1$. Thus, $\cS_1$ has
infinitely many extreme points, and so it is not a polytope.

The set of extreme points of any face $F$ of $\cS_1$, is just the set $F\cap R_\cH$ of all pure product states contained in $F$.
It is not known what is the maximal dimension of the proper
faces $F$ of $\cS_1$.

{\em Observation 2}. If $\r\in\cS$ has deficient rank then $\r\in\pars$. This follows from the fact that $\r\in\pard$.

In particular, the condition is satisfied if the range of
$\r\in\cS$ contains only finitely many product vectors. (We
always count the product vectors up to a scalar factor.)

{\em Observation 3}. The set $R_V$ is the set of extreme points of $F_V$. Consequently, $F_V$ is the convex hull of $R_V$.

 \bpf
Every extreme point $\r$ of $F_V$ is also an extreme point of
$\cS_1$ and so $\r=\proj{x}$ for some unit product vector
$\ket{x}$. Since $\r\in\tilde{F}_V$ we have $\ket{x}\in V$.
Thus $\ket{x}\in P_V$ and so $\r\in R_V$. The converse is
obvious.
 \epf

Very little is known about the set $\cF$ of all faces of $\cS_1$.
Some proper faces of low dimensions have been explicitly
constructed. Most of them are polytopes. Examples of proper
faces that are not polytopes can be found in \cite{as10}. Such examples exist even in the case of two qubits $(n=d_1=d_2=2)$.
Indeed, the face associated to $\ket{0}\otimes\cH_2$ is
$\proj{0}\otimes B_2$ where $B_2$ is the Bloch ball of the second
qubit. By using Observation 3, we can construct a rich family of
faces of $\cS_1$ which are polytopes. Under a suitable condition
on the $d_i$, these faces are not simplices.

 \bex \label{ex:Pi} {\rm
A generic subspace $V\subset\cH$ of dimension $d-\sum(d_i-1)$
contains exactly
 \bea
N:=\frac{\left(\sum(d_i-1)\right)!}{\prod\left((d_i-1)!\right)}
 \eea
product vectors. First, the number $N$, given by this formula, is the degree of the product of complex projective spaces associated to the $\cH_i$, $i=1,\ldots,n$, under the Segre
embedding into the projective space associated to $\cH$
(see \cite[p. 412]{lan12}). Second, the claim that $N$ is also  the number of product vectors (up to a scalar factor) in the 
generic subspace of dimension $d-\sum(d_i-1)$ follows immediately 
from one of the definitions od the degree of irreducible 
projective varieties, 
see \cite[Definition 18.1, (iii)]{Harris:1992}.

Let $\ket{z_i}\in P_V$,
$i=1,\ldots,N$, be pairwise non-parallel. Then the convex hull
$\Pi$ of the states $\proj{z_i}$ is a face of $\cS_1$. Each of the states $\proj{z_i}$ is an extreme point of $\Pi$, and $\Pi$
is not a simplex if $N\ge d$. Since the $\ket{z_i}$ span $V$
(see \cite{cd13CMP}), we have $\dim\Pi\ge d-\sum(d_i-1)$.
This example also shows that the converse of Observation 3
is not valid, e.g., when $N\ge d$. We leave as an open problem the computation of the dimension of $\Pi$.
 \qed }\eex

\textbf{Open problem 1}. \quad Compute the dimension of the polytope $\Pi$.

Let us remark that if $\r\in\cS$ then in fact $\r^\G\in\cS$ for all $\G\in\T$. Since each $\G\in\T$ is an invertible linear
transformation of $H$ which preserves $\cS_1$, it must map  faces to faces, preserve their dimensions and other properties of faces such as being exposed, maximal, induced etc.
We say that a Hermitian operator $\r\in H$ is
{\em full} if  $\r^\G$ has full rank for all $\G\in\T$. 

Observation 2 can be generalized as follows.
 \bpp
 \label{pp:ParTransp}
If a state $\r\in\cS$ is not full, then $\r\in\pars$.
 \epp
 \bpf
By the hypothesis, there exists $\G\in\T$ such that
$\rank\r^\G<d$. By the above remark $\r^\G\in\cS$ and
Observation 2 implies that $\r^\G\in\pars$.
As $\G(\pars)=\pars$, we conclude that $\r\in\pars$.
 \epf

\subsection{Full states on $\pars$}

We say that an entangled PPT state is a PPTES. Note that the set of all full states $\r\in H$ is open in $H$, and that all states $\r$ in the interior of $\cS$ are full. If $\r\in\pars$ is a full state then, for sufficiently small $t>0$, $\r-t I_d$ is a PPTES. If $d\le6$ there are no PPTES and so, in these cases there are no full states on $\pars$. In particular, two-qubit proper faces of $\cS_1$ do not contain any full state. We shall see in
Proposition \ref{pp:NonTrivial} below that this is not true if
$d>6$. For a concrete example in $3\otimes3$ see Example
\ref{ex:FinF'} below.

In quantum information, the state $I_d/d$ is regarded as the white noise to the initial normalized PPTES $\r$. There is a
unique $p\in(0,1)$ such that $\r(t):=tI_d/d+(1-t)\r$ is entangled
for $t\in[0,p)$ and $\r(p)\in\pars_1$.
The bigger $p$ is, the more robust entanglement of $\r$ is
against the decoherence. For a given $\r$, it is an important
question to analytically compute $p$. It shows how quantum correlation of $\r$ is removed by decoherence with the
environment. However it is usually hard to compute $p$, as
there are very few tools to decide whether a PPT state is separable (except in the two-qubit and qubit-qutrit cases \cite{horodecki97}).
In contrast, starting with a full state $\s\in\pars_1$, we may
choose $\r=(1+t)\s-tI_d/d$ with small $t>0$ in which case we have have $p=t/(1+t)$. It gives us a method of analytically
deciding the robustness of a PPTES against the noise. So it is
a meaningful problem to construct full states on $\pars_1$.

For a long time no explicit full states on $\pars_1$ in any
quantum systems that we consider were known. As observed above,
if $d\le6$ there are no such states. Two well known examples of bipartite states with $d_1=d_2$ are the Werner state \cite{werner89}
 \bea \label{eq:werner}
I_d -\frac{1}{d_1}
\sum^{d_1-1}_{i,j=0}\ketbra{ij}{ji},
 \eea
and the isotropic state \cite{hh99}
 \bea \label{eq:isotropic}
I_d +\sum^{d_1-1}_{i,j=0}\ketbra{ii}{jj}.
 \eea
A family of separable states of full rank on $\pars$ in $2\ox d_2$ has been constructed in \cite[Proposition 5]{cd12PRA}. It is
also known (see \cite{dc00}) that the multiqubit state
 \bea
 I_{2^n} +\proj{\ghz},\quad
(\ket{\ghz}=\ket{0,\ldots,0}+\ket{1,\ldots,1})
 \eea
is separable. Although all of these states have full rank and lie on $\pars$, it turns out that none of them is a full state.

 \bpp
 \label{pp:NonTrivial}
If a proper face $F$ of $\cS_1$ contains a full state, then $F$ is not induced. If $d>6$ then there exist full states on $\pars$
(and $\pars_1$).
 \epp
 \bpf
Let $\r\in F$ be a full state. Then $\r^\G$ belongs to the interior of $\cD_1$ for all $\G\in\T$. Consequently, $F$ is not induced. To prove the second assertion, we assume that $d>6$.
Then there exist PPTES and we fix one of them, say $\r$. 
The line segment joining $I_d$ to $\r$ contains a unique point 
$\s\in\pars$. Since $\r^\G\ge0$ for all $\G\in\T$, it follows
that $\s^\G$ has full rank for all $\G\in\T$. Thus $\s\in\pars$ 
is a full state.
 \epf

We point out that the converse of the first assertion is false 
(see the discussion below Proposition \ref{pp:maxfacedim}).

The first explicit examples of full states on $\pars_1$ have been
constructed recently in \cite[p. 18]{hk14}, see the example below. (Our terminology has not been used in that paper.) We shall analyze this family in more details in section \ref{sec:3x3}.

 \bex \label{ex:FinF'} {\rm
Let us consider the bipartite system $3\otimes3$. In the paper
\cite{hk14} the authors have constructed 10 normalized real product vectors $\ket{z_i}$, $i=1,\ldots,10$, depending on one real parameter $b>0$, $b\ne1$. It is easy to verify that any 9 of them are linearly independent. They have shown that the convex hull, $\D_b$, of the 10 states $\proj{z_i}$ is a 9-dimensional simplex and that $\D_b\in\cF$. Thus $\cR(\D_b)=\cH$ and $\D_b\su F_{\cH}=\cS_1$. Since all vectors $\ket{z_i}$ are real, it follows that $\r^{\G_1}=\r$. Hence, each interior point $\r$ of $\D_b$ is a full state. By Proposition \ref{pp:NonTrivial}, the face $\D_b$ (as well as any proper face containing it) is not induced. 
 \qed }\eex

The full states on $\pars_1$ in the above example have unique decomposition as the convex sum of pure product states. This is not surprising because separable states with a unique decomposition evidently belong to $\pars_1$. However the converse
is not true. For example, for $n=2$ any state $\ketbra{0}{0}\ox 
\s$, where $\s$ is a mixed state, has infinitely many decompositions as the convex sum of pure product states.

In contrast to Example \ref{ex:FinF'}, we claim that in the $2\ox4$ system there is no full state $\r\in\pars$ such that $\r^{\G_1}=\r$. Suppose there is such a state $\r$. Then $\s:=\r-t I_8$ is a PPTES for small $t>0$. As $\s^{\G_1}=\s$, this contradicts \cite[Theorem 2]{kck00}. However, we do not know whether in this system there is a full state $\r\in\pars$ such that $\r^{\G_2}=\r$.

\textbf{Open problem 2}. \quad Construct a concrete example of a
full state on $\pars$ in the $2\otimes2\otimes2$ system.

The following lemma shows that suitable tensor products are
full separable states lying on $\pars$. We define the tensor
product of two $n$-partite states, $\r_{A_1,\ldots,A_n}$ acting
on $\cH_A$ and $\s_{B_1,\ldots,B_n}$ acting on $\cH_B$, as a new
$n$-partite state $(\r\ox\s)_{C_1,\ldots,C_n}$ acting on $\cH_C$
where each system $C_i$ is obtained by combining $A_i$ and $B_i$ into one system. We shall use subscripts $A,B,C$ to distinguish
these three $n$-partite systems. E.g., $I_C$ is the identity operator on $\cH_C$,  and $\cS_C$ is the cone of non-normalized
separable states in $\cH_C$, etc.

 \bl
 \label{le:genericpoint=prod}
Let $\r\in\cS_A$ and $\s\in\cS_B$ be full states normalized so that $\tr(\r-I_A)=\tr(\s-I_B)=0$. Then $\a:=\r\ox\s$ is a full
separable state. Moreover, $\a\in\pars_C$ if and only if
$\r\in\pars_A$ or $\s\in\pars_B$.
 \el
 \bpf
The first assertion is obvious. To prove the second assertion,
assume that $\a$ is an interior point of $\cS_C$. Then
$\a- t I_C\in\cS_C$ for small $t>0$. By tracing out the systems $B_1,\ldots,B_n$, we see that $\r-t I_A\in\cS_A$ for small $t>0$. Thus, $\r$ must belong to the interior of $\cS_A$. Similarly,
$\s$ belongs to the interior of $\cS_B$.
Conversely, assume that $\r$ and $\s$ are interior points of $\cS_A$ and $\cS_B$, respectively. Then there is a $t>0$ such
that $\r-tI_A\in\cS_A$ and $\s-tI_B\in\cS_B$. It follows easily
that $\a-t^2I_C\in\cS_C$. Hence, $\a$ is an interior point of
$\cS_C$.
 \epf

Let $\r\in\pars_A$ be any full two-qutrit state mentioned in  Example \ref{ex:FinF'}. It follows from Lemma \ref{le:genericpoint=prod} that the bipartite separable state $\r\ox I_{s_1s_2}$ is a full state on $\pars$ on the space $(\bC^3\ox\bC^{s_1})\ox(\bC^3\ox\bC^{s_2})$ for any positive integers $s_1,s_2$. 

So far we have discussed the full separable states of bipartite
systems. Here we construct some full states on $\pars$ in
multipartite systems. Let $\r$ and $\s$ be $l$ and $m$-partite
separable states, respectively. We regard $\b=\r\ox\s$ as a
$(l+m)$-partite separable state. If $\r$ or $\s$ is a full
state on  $\pars$, then so is $\b$. This assertion follows from the fact that a state $\b-tI_d$ is entangled for $t>0$. Let $\r$ be the two-qutrit full state on $\pars$ constructed in Example \ref{ex:FinF'}, so $l=2$. By choosing a separable state $\s$, with $m$ arbitrary, we obtain a full state $\b$ on $\pars$ in a  multipartite system.

To conclude this section, we present the following observation
concerning the full states on $\pars$.
 \bl
 \label{le:sep+ent=ent}
Suppose $\r\in\pars$ is a full state. Then there exists
$\ket{\ps}\in\cH$ such that $p\r+\proj{\ps}$ is entangled for all
$p\ge0$.
 \el
 \bpf
Since $\r$ has full rank, there exists $t>0$ such that
$\r-tI_d>0$. Thus, we have $\r-tI_d=\sum_i \proj{\ps_i}$ where
the sum is finite. Assume that for each $i$ there is a $p_i\ge0$ such that $\r_i:=p_i\r+\proj{\ps_i}\in\cS$. The identity
$\r-stI_d=(1-s(1+\sum p_i))\r+s\sum_i \r_i$ is valid for all
real $s$. Consequently, for small $s>0$ we have $\r-stI_d>0$
which contradicts the fact that $\r\in\pars$. Hence, for at least
one index $i$ the state $p\r+\proj{\ps_i}$ must be entangled
for all $p\ge0$.
 \epf

The lemma implies that some entangled state $\ket{\ps}$ may be
``eternally" robust to some separable state $\r$, which is regarded
as noise in quantum information. When $p$ is large, the entangled
state $p \r + \proj{\ps}$ will become a PPTES.

Let us say that a point $\s\in\pars_1$ is a {\em smooth point}
of $\pars_1$ if the intersection of $\pars_1$ with a small
ball $B_\varepsilon:=\{ \r\in H:\|\r-\s\|<\varepsilon \}$ is a
smooth manifold. In connection with the Example \ref{ex:Pi}
we ask whether the full states belonging to the polytope $\Pi$
are smooth.

\section{Some faces of $\cS_1$ when $d_1=\cdots=d_n$}
\label{sec:Some}

The following lemma and its corollary will be used in several subsequent proofs.
 \bl \label{le:Monomials}
Let $z_1,\ldots,z_m$ be independent complex variables and $n$ a positive integer. Then the monomials
$z_1^{j_1}(z_1^*)^{k_1}\cdots z_m^{j_m}(z_m^*)^{k_m}$, where 
$j_1,k_1,\ldots,j_m,k_m$ are nonnegative integers, are linearly 
independent over complex numbers. More precisely, if 
$P:=P(x_1,y_1,\ldots,x_m,y_m)$ is a polynomial with complex coefficients in $2m$ independent commuting variables 
$x_1,y_1,\ldots,x_m,y_m$ and $P(z_1,z_1^*,\ldots,z_m,z_m^*)$
is identically zero, then $P=0$, i.e., all coefficients of 
$P$ are zero.
 \el
 \bpf
We use induction on $m$. The assertion is obviously valid when $m=1$, i.e., $z_1$ and $z_1^*$ are algebraically independent over $\bC$. Assume that $m>1$. We have $P=\sum_{j,k} x_m^j y_m^k P_{j,k}(x_1,y_1,\ldots,x_{m-1},y_{m-1})$, where $P_{j,k}$ are polynomials in the $2(m-1)$ variables. By the hypothesis of the lemma we have the identity 
$$\sum_{j,k} z_m^j (z_m^*)^k 
P_{j,k}(z_1,z_1^*,\ldots,z_{m-1},z_{m-1}^*) =0.$$ 
Since $z_1$ and $z_1^*$ are algebraically independent over $\bC$, 
we deduce that each coefficient 
$P_{j,k}(z_1,z_1^*,\ldots,z_{m-1},z_{m-1}^*)$ is identically zero. By the induction hypothesis, we conclude that each  polynomial $P_{j,k}(x_1,y_1,\ldots,x_{m-1},y_{m-1})$ is zero. Consequently, $P=0$.
\epf

The following corollary is an easy consequence of the lemma.
 \bcr \label{cr:MonDim}
Let $\mu_1,\mu_2,\ldots,\m_s$ be distinct monomials in the complex variables $z_1,\ldots,z_m$ and their conjugates 
$z_1^*,\ldots,z_m^*$. Assume that this list of monomials contains exactly $a$ real-valued monomials, exactly $b$ pairs $\{\mu,\mu^*\}$, $(\mu^*\ne\mu)$, of complex-conjugate monomials, and $c$ additional complex-valued monomials. (Thus $s=a+2b+c$.) 
Let $V$ be a complex vector space and $v_1,\ldots,v_s$ linearly 
independent vectors of $V$. If $L$ is the real span of the set 
of vectors $\{\sum_{k=1}^s \m_k v_k: z_1,\ldots,z_m\in\bC\}$, 
then $\dim L=a+2(b+c)$.
 \ecr

We can write any linear operator $L$ on $\cH$ as
 \bea \label{eq:OperL}
L=\sum L_{k_1,\ldots,k_n}^{j_1,\ldots,j_n}
\ketbra{j_1,\ldots,j_n}{k_1,\ldots,k_n},
 \eea
where the summation is over all
$j_i,k_i\in\{0,1,\ldots,d_i-1\}$, $i=1,\ldots,n$, and the
components (we refer to them also as ``matrix coefficients")
are given by $L_{k_1,\ldots,k_n}^{j_1,\ldots,j_n}=
\bra{j_1,\ldots,j_n}L\ket{k_1,\ldots,k_n}$. 

In this section we assume that $\cH_1=\cdots=\cH_n$ and so
$\cH=\otimes^n \cH_1$. We denote by $\cH_{\rm sym}$ the subspace of $\cH$ consisting of symmetric tensors. We say that $L$ is {\em range-symmetric} if $\cR(L)\sue\cH_{\rm sym}$. We warn the reader that the matrix of a range-symmetric operator is not necessarily symmetric. One can recognize whether $L$ is range-symmetric by examining its matrix coefficients. This is the case if and only if the $L_{k_1,\ldots,k_n}^{j_1,\ldots,j_n}$ are invariant under the permutations of the superscripts $j_1,\ldots,j_n$. If $L$ is Hermitian, this implies that the
$L_{k_1,\ldots,k_n}^{j_1,\ldots,j_n}$ are also invariant under
the permutations of the subscripts $k_1,\ldots,k_n$.
We shall denote by $H_{\rm s}$ the subspace of $H$ consisting of
all range-symmetric operators,
 \bea \label{eq:RangeSym}
H_{\rm s}=\{\r\in H:\cR(\r)\sue\cH_{\rm sym}\}.
 \eea
Since the space $\cH_{\rm sym}$ has (complex) dimension $\binom{n+d_1-1}{n}$, we have
 \bea \label{eq:dimHrs}
\dim H_{\rm s} &=& \binom{n+d_1-1}{n}^2.
 \eea

\subsection{The face $F_{\cH_{\rm sym}}$}
\label{subsec:TheFace}

We consider here the induced face $F_{\cH_{\rm sym}}$ of $\cS_1$. 
In view of \eqref{eq:RangeSym} we have 
 \bea \label{eq:F-Sym}
F_{\cH_{\rm sym}}=H_{\rm s}\cap\cS_1.
 \eea
We shall describe the set of extreme points of this face and compute its dimension.

For any nonzero $W\in H$ we denote by $X_W$ the hyperplane of
$H$ defined by the equation $\tr(W\r)=0$, and we set
$F_W=X_W\cap\cS_1$. If $\tr(W\r)\ge0$ for all $\r\in\cS_1$, then $F_W$ is an exposed proper face of $\cS_1$ (possibly empty). 
This face $F_W$ should not be confused with the previously defined face $F_V$ (see \eqref{eq:FaceF_V}). It will be always clear from the context whether the subscript of $F$ is a subspace of $\cH$ or a nonzero Hermitian operator on $\cH$.

 \bpp \label{pp:SymOper}
Let $W_{\rm s}=I_d-P_{\rm s}$, where $P_{\rm s}\in H_{\rm s}$ is the projector onto $\cH_{\rm sym}$, and let $F$ be the convex hull of the set of operators $\proj{x,\ldots,x}\in H$ with $\ket{x}\in\cH_1$ and $\|x\|=1$.

(i) For any product vector $\ket{x_1,\ldots,x_n}\in\cH$, we have
$\bra{x_1,\ldots,x_n}W_{\rm s}\ket{x_1,\ldots,x_n}\ge0$ and the
equality holds if and only if all $\ket{x_i}$ are parallel to each other. Moreover, $F=F_{\cH_{\rm sym}}$.

(ii) The subspace $H_{\rm s}$ is spanned by $F$ and
 \bea \label{eq:dimF}
\dim F &=& \binom{n+d_1-1}{n}^2 -1.
 \eea
 \epp
 \bpf
(i) We may assume that $\|x_i\|=1$ for each $i$. Then 
$\bra{x_1,\ldots,x_n} W_{\rm s} \ket{x_1,\ldots,x_n}=
1-\|P_{\rm s}\ket{x_1,\ldots,x_n}\|^2\ge0$ and the equality holds if and only if $\ket{x_1,\ldots,x_n}\in\cH_{\rm sym}$, i.e.,
if and only if the $\ket{x_i}$ are parallel to each other.
It follows that $F=X_{W_{\rm s}}\cap\cS_1$. As $H_{\rm s}\sue 
X_{W_{\rm s}}$, we have $F_{\cH_{\rm sym}}\sue F$. The opposite 
inclusion is immediate from the definition of $F$. Hence, we have 
$F=F_{\cH_{\rm sym}}$.

(ii) To prove the first assertion of (ii), note that
$\proj{x,\ldots,x}\in H_{\rm s}$ for all $\ket{x}\in\cH_1$.
Assume that the assertion is false. Then there
exists a nonzero $L\in H_{\rm s}$ such that
$\bra{x,\ldots,x}L\ket{x,\ldots,x}=0$ for all $\ket{x}\in\cH_1$. By
using the expansion $\ket{x}=\sum_{j=0}^{d_1-1} \xi_j\ket{j}$, we
obtain that
 \bea \label{eq:JedL}
\sum L_{k_1,\ldots,k_n}^{j_1,\ldots,j_n}
\xi_{j_1}^* \cdots \xi_{j_n}^* \xi_{k_1}\cdots\xi_{k_n}=0,
 \eea
where the summation is over all pairs of repeated indexes,
each index running through the integers $0,1,\ldots,d_1-1$.
Since $L$ is range-symmetric and Hermitian, the components
$L_{k_1,\ldots,k_n}^{j_1,\ldots,j_n}$ are invariant under
the permutation of the subscripts or superscripts. By collecting
the like terms in \eqref{eq:JedL}, we obtain the identity
 \bea
\sum_{\begin{array}{c}
0\le j_1\le\cdots\le j_n<d_1\\
0\le k_1\le\cdots\le k_n<d_1 \end{array}}
\mu_{k_1,\ldots,k_n}^{j_1,\ldots,j_n}
L_{k_1,\ldots,k_n}^{j_1,\ldots,j_n}
\xi_{j_1}^* \cdots \xi_{j_n}^* \xi_{k_1}\cdots\xi_{k_n}=0,
 \eea
where $\mu_{k_1,\ldots,k_n}^{j_1,\ldots,j_n}$ are some positive
integers. By Lemma \ref{le:Monomials}, all components
$L_{k_1,\ldots,k_n}^{j_1,\ldots,j_n}$ must vanish. Thus
$L=0$ and we have a contradiction.

The second assertion follows from \eqref{eq:dimHrs} by taking
into account that $F\sue H_1$.
 \epf

For any $\G\in\T$ the image
$\G(H_{\rm s}\cap\cS_1)=\G(H_{\rm s})\cap\cS_1$ is also a face
of $\cS_1$ having the same dimension as $H_{\rm s}\cap\cS_1$.
Consequently, the following corollary is valid.
 \bcr \label{cr:F^Gamma}
Let $S$ be any subset of $\{1,\ldots,n\}$ and $\G_S=\prod_{i\in S} \G_i$. For any $\ket{x}\in\cH_1$ we set $\ket{x_S}:=\ket{x_1,\ldots,x_n}$ where
$\ket{x_i}=\ket{x^*}$ for $i\in S$ and $\ket{x_i}=\ket{x}$
otherwise. Then the face $\G_S(H_{\rm s})\cap\cS_1$ is the
convex hull of all $\proj{x_S}$ where $\ket{x}$ runs over all
unit vectors in $\cH_1$.
 \ecr

Let $H^{\rm re}$ denote the subspace of $H$ consisting of the
operators $L$ such that all matrix coefficients of $L$
are real. We remark that the dimension of the subspace
$H_{\rm s}^{\rm re}:=H^{\rm re}\cap H_{\rm s}$ is given by
 \bea \label{eq:dimHres}
\dim H_{\rm s}^{\rm re} &=& \frac{1}{2} \binom{n+d_1-1}{n}
\left[ \binom{n+d_1-1}{n}+1 \right].
 \eea
To prove this formula, let us denote by $\cH_1^{\rm re}$
the real Hilbert space consisting of all
$\ket{x}=\sum_{j=0}^{d_1-1} \xi_j\ket{j}$ with all $\xi_j$ real,
and let $\cH^{\rm re}=\otimes^n_\bR\cH_1^{\rm re}$ be the real subspace of $\cH$ consisting of the tensors having all
components real. Then the space $H_{\rm s}^{\rm re}$ can be
identified with the space of symmetric and range-symmetric
operators on $\cH^{\rm re}$.
Now the formula \eqref{eq:dimHres} follows from the fact that
the space of the symmetric tensors in $\cH^{\rm re}$ has
dimension $\binom{n+d_1-1}{n}$.

Let $H^\Theta$ denote the subspace of $H$ consisting of all
operators $L$ fixed under $\T$, i.e.,
 \bea \label{eq:H^Theta}
H^\T &=& \{L\in H: L^\G=L,~\forall\G\in\T\}.
 \eea
Its dimension was computed in general (for arbitrary
$d_1,\ldots,d_n$) in \cite{cd13JMP} where it was also observed
that $H^\T\sue H^{\rm re}$.
Finally, we set $H_{\rm s}^\T=H^\T\cap H_{\rm s}$.

Let $\L_m$ be the set of all integer sequences
$l=(l_1,l_2,\ldots,l_m)$ such that 
$0\le l_1\le l_2\le\ldots\le l_m<d_1$. 
and $l_i\in\{0,1,\ldots,d_1-1\}$ for all $i$.
We shall write $(j_1,\ldots,j_m)\to(l_1,\ldots,l_m)$ if
$(l_1,\ldots,l_m)\in\L_m$ and there is a permutation $\s$
of $\{1,\ldots,m\}$ such that $j_{\s i}=l_i$ for all $i$.

For each $l\in\L_{2n}$ let
 \bea \label{eq:BazTilda}
\r[l]:=
\sum_{(j_1,\ldots,j_n,k_1,\ldots,k_n)\to l}
\ketbra {j_1,\ldots,j_n} {k_1,\ldots,k_n}.
 \eea
It is easy to verify that $\r[l]\in H_{\rm s}^\T$ for
all $l\in\L_{2n}$. We claim that the set
$\{\r[l]:l\in\L_{2n}\}$ is a basis of
$H_{\rm s}^\T$. It is obvious that this is an orthogonal set of
vectors in $H_{\rm s}^\T$. Let $L\in H_{\rm s}^\T$ be arbitrary.
We can write it as in \eqref{eq:OperL}. Since
$H_{\rm s}^\T\sue H^{\rm re}$, all components
$L_{k_1,\ldots,k_n}^{j_1,\ldots,j_n}$ are real. Moreover, we
know that if
$(j_1,\ldots,j_n,k_1,\ldots,k_n)\to l$,
then $L_{k_1,\ldots,k_n}^{j_1,\ldots,j_n}=
L^{l_1,\ldots,l_n}_{l_{n+1},\ldots,l_{2n}}$.
This means that $L$ is a real linear combination of the set
$\{\r[l]:l\in\L_{2n}\}$ and our claim is
proved. Consequently,
 \bea \label{eq:dimHsT}
\dim H_{\rm s}^\T &=& \binom{2n+d_1-1}{2n}.
 \eea

\subsection{The intersection of all 
$\G(H_{\rm s})\cap\cS_1$, $\G\in\T$ }

Since $H_{\rm s}\cap\cS_1$ is a face of $\cS_1$, the same is true
for its image $\G(H_{\rm s})\cap\cS_1$ under $\G\in\T$. Our
objective here is to determine the intersection of all these
faces. For that purpose we need the following proposition.

 \bpp \label{pp:Presek}
We have $H^\T_{\rm s}=\cap_{\G\in\T} \G(H_{\rm s})=
H_{\rm s}\cap \G_1(H_{\rm s})$.
 \epp
 \bpf
Obviously, we have $H^\T_{\rm s}\sue\cap_{\G\in\T}\G(H_{\rm s})
\sue H_{\rm s}\cap \G_1(H_{\rm s})$. Hence, it suffices to show that $H_{\rm s}\cap\G_1(H_{\rm s})\sue H^\T_{\rm s}$.
Let $L\in H_{\rm s}\cap\G_1(H_{\rm s})$ be arbitrary and
write it as in \eqref{eq:OperL}. We have to show that
$L\in H^\T$, i.e., that the components
$L_{k_1,\ldots,k_n}^{j_1,\ldots,j_n}$
remain unchanged when we permute arbitrarily the $2n$ indexes
$j_1,\ldots,j_n,k_1,\ldots,k_n$. Since $L\in H_{\rm s}$, we
know that these components do not change when we permute
the superscipts and subscripts separately. As
 \bea \label{eq:CompL^G}
L^{\G_1}=\sum L_{j_1,k_2,\ldots,k_n}^{k_1,j_2,\ldots,j_n}
\ketbra{j_1,j_2,\ldots,j_n}{k_1,k_2,\ldots,k_n}
\in H_{\rm s},
 \eea
the components $L_{j_1,k_2,\ldots,k_n}^{k_1,j_2,\ldots,j_n}$
are unchanged when we permute the indexes $j_1,j_2,\ldots,j_n$.
Equivalently, the components
$L_{k_1,k_2,\ldots,k_n}^{j_1,j_2,\ldots,j_n}$ are unchanged when we permute the indexes $k_1,j_2,j_3,\ldots,j_n$.

For convenience, let us label the superscripts $j_1,\ldots,j_n$ and the subscripts $k_1,\ldots,k_n$ of
$L_{k_1,k_2,\ldots,k_n}^{j_1,j_2,\ldots,j_n}$ with integers
$1,\ldots,n$ and $n+1,\ldots,2n$, respectively.
The symmetric group $S_{2n}$ permutes the set
$\Omega:=\{1,2,\ldots,2n\}$. We single out the three subgroups of
$S_{2n}$, each isomorphic to $S_n$: the first one permutes
only the integers $1,\ldots,n$, the second one permutes only the
integers $n+1,\ldots,2n$, and the third permutes only the
integers  $2,3,\ldots,n,n+1$. We have shown above that the
components $L_{k_1,\ldots,k_n}^{j_1,\ldots,j_n}$ are not changed
when we permute the indexes by using a permutation belonging
to one of these three copies of $S_n$.
Now our assertion follows from the fact that these three copies
of $S_n$ generate the whole group $S_{2n}$.

To prove this fact, let us denote by $G$ the subgroup of $S_{2n}$
generated by our three copies of $S_n$. It is obvious that $G$
acts transitively on the set $\Omega$. It is also obvious that
$G$ is primitive, i.e., there is no proper subset $\D$ of
$\Omega$ of cardinality at least 2 such that for each $g\in G$
either $g(\D)=\D$ or $g(\D)\cap\D=\emptyset$. Since $G$ contains
a transposition, we must have $G=S_{2n}$
(see e.g., \cite[Chap. II, Satz 4.5]{bh67}).
  \epf

As an example, we mention that in the case $n=2$, $d_1=d_2=3$, the spaces $H$, $H^{\rm re}$, $H_{\rm s}$, $H^\T$ and $H^\T_{\rm s}$ have dimensions 81,45,36,36 and 15, respectively.

 \bpp \label{pp:RealVersion}
Let $F$ be the face $F_{\cH_{\rm sym}}$ of $\cS_1$ and let 
$F':=\cap_{\G\in\T}\G(F)$.

(i) The operators $\proj{x,\ldots,x}\in H$ with
$\ket{x}\in\cH_1^{\rm re}$ span the space $H_{\rm s}^\T$.

(ii) The face $F'$ is the convex hull of all $\proj{x,\ldots,x}\in H$ with $\ket{x}\in\cH_1^{\rm re}$ and $\|x\|=1$.

(iii) $F'$ is neither induced nor a polytope.
 \epp
 \bpf
(i) Note first that $\proj{x,\ldots,x}\in H_{\rm s}^\T$ for all
$\ket{x}\in\cH_1^{\rm re}$. Assume that the assertion is false. Then
there exists a nonzero $L\in H_{\rm s}^\T$ such that
$\bra{x,\ldots,x}L\ket{x,\ldots,x}=0$ for all $\ket{x}\in\cH_1^{\rm
re}$. By using the expansion $\ket{x}=\sum_{j=0}^{d_1-1}
\xi_j\ket{j}$, we obtain that
 \bea \label{eq:JedLR}
\sum L_{k_1,\ldots,k_n}^{j_1,\ldots,j_n}
\xi_{j_1} \cdots \xi_{j_n} \xi_{k_1}\cdots\xi_{k_n}=0,
 \eea
where the summation is over all pairs of repeated indexes,
each index running through the integers $0,1,\ldots,d_1-1$.
Recall that the components
$L_{k_1,\ldots,k_n}^{j_1,\ldots,j_n}$ are symmetric in the
subscripts and the superscripts. Moreover, since
$L\in H^\T$, these components are not changed if we switch
one of the subscripts with one of the superscripts. Hence,
by collecting the like terms in \eqref{eq:JedL}, we obtain
an identity
 \bea
\sum_{1\le m_1\le\cdots\le m_{2n}\le 2n}
\mu'_{m_1,\ldots,m_{2n}}
L^{m_1,\ldots,m_n}_{m_{n+1},\ldots,m_{2n}}
\xi_{m_1} \cdots \xi_{m_{2n}}=0,
 \eea
where $\mu'_{m_1,\ldots,m_{2n}}$ are some positive integers.
Since the monomials $\xi_{m_1} \cdots \xi_{m_{2n}}$,
$1\le m_1\le\cdots\le m_{2n}\le 2n$, are linearly independent,
all components $L_{k_1,\ldots,k_n}^{j_1,\ldots,j_n}$ must vanish.
Thus $L=0$ and we have a contradiction.

(ii) By Proposition \ref{pp:Presek} we have
$$
F' = \cap_{\G\in\T} (\G(H_{\rm s})\cap\cS_1)
=\left( \cap_{\G\in\T} \G(H_{\rm s}) \right) \cap\cS_1
=H_{\rm s}^\T\cap\cS_1.
$$
Note that $H_{\rm s}^\T\cap\cS_1$ is a face of $F$. Hence, 
every extreme point $\r$ of $H_{\rm s}^\T\cap\cS_1$ is also an
extreme point of $F$. By Proposition \ref{pp:SymOper} we have $\r=\proj{x,\ldots,x}$ for some unit vector $\ket{x}\in\cH_1$. Since $\r\in H_{\rm s}^\T$, it follows that $\r^{\G_1}=\r$ 
and so $\ketbra{x^*}{x^*}=\ketbra{x}{x}$. Thus, up to a phase factor, $\ket{x}\in\cH_1^{\rm re}$ and the assertion is proved.

(iii) Assume that $F'$ is induced, i.e., $F'=\G(F_V)$ for 
some $\G\in\T$ and some subspace $V\sue\cH$. By (ii) $P_{F'}$ consists of real product vectors (up to a phase factor). It follows that $F'=F_V$. By \eqref{eq:P_{F_V}} we have 
$P_{F'}=P_{F_V}=P_V$. Hence 
$\cH_{\rm sym}=\cR(F')=\lin P_V\sue V$. Consequently, 
$P_{\cH_{\rm sym}}\sue P_V=P_{F'}$. If $\ket{x}\in\cH_1$ is a unit vector which is not real (up to a phase factor), then
$\proj{x,x}\in P_{\cH_{\rm sym}}\setminus P_{F'}$ and we have a contradiction. Thus, $F'$ is not induced. As the set of extreme points of $F'$ is infinite, $F'$ is not a polytope.
 \epf

We remark that the face $F'$ is exposed because by its definition it is the intersection of exposed faces.

\section{Induced maximal faces of $\cS_1$}
\label{sec:InducedMax}

The problem of describing the proper faces of $\cS_1$ can be split into two steps: first describe all maximal faces and second describe the proper faces of the maximal faces. To simplify this problem further, we can use the group $\GL:=\times_{i=1}^n\GL(d_i)$ which acts on $\cH$ as the group of invertible local operators (ILO). The linear action of $A\in\GL$ sends $\ket{x}\in\cH$ to $A\ket{x}$. By the induced action on $H$, $A$ sends $\r\in H$ to $A\r A^\dag$. The former action does not preserve the norm of vectors, and the latter does not preserve the trace of the Hermitian operators. Hence, we are forced to use the actions where $A$ sends $\ket{x}\to A\diamond\ket{x}:=A\ket{x}/\|Ax\|$ and sends $\r\to A\diamond\r:=(A\r A^\dag)/\tr(A\r A^\dag)$. The $\diamond$-action is well defined on the set $\cD_1$. One can easily verify that it maps $\cS_1$ onto itself, and preserves the convexity property. Moreover, it permutes the faces of $\cS_1$, preserves their dimensions, as well as the properties of being maximal or exposed. Instead of classifying the faces of $\cS_1$ up to the action of the local unitary group, we shall consider the easier problem of classifying the faces up to the above  $\diamond$-action of $\GL$. We say that two faces $F,F'\in\cF$ are {\em $\GL$-equivalent} if $F'=A\diamond F$ for some $A\in\GL$.

For any subspace $V\sue\cH$ we have
 \bea
A\diamond R_V=\{ \frac{A\proj{v}A^\dag}{\|Av\|^2} :
\ket{v}\in P_V \}=R_{AV},
 \eea
and consequently
 \bea \label{eq:F_AV}
A\diamond F_V=F_{AV},\quad A\in\GL.
 \eea
It follows that the $\diamond$-action maps induced faces to induced faces.

If a subspace $V\sue\cH$ has dimension $<d-1$, then it is easy to see that the face $F_V$ is proper but not maximal. Indeed, there exist
$\ket{v}\in P_\cH \setminus V$ and so if $V'=V+\bC\ket{v}$ then
$F_V\subset F_{V'}\subset\cS_1$. 

 \bpp
 \label{pp:maxfacedim}
If $F$ is a maximal face of $\cS_1$, then $\dim\cR(F)\ge d-1$.
Consequently, a maximal face $F\in\cF$ is induced if and only if $\dim\cR(\G(F))=d-1$ for some $\G\in\T$.
 \epp
 \bpf
Let $F\in\cF$ be such that $\dim\cR(F)<d-1$. Choose a hyperplane 
$V$ of $\cH$ such that $\cR(F)\su V$. Then $F\sue F_{\cR(F)}\su 
F_V\su\cS_1$ and so $F$ is not maximal. The first assertion 
follows. Next we prove the second assertion. The necessity of the
condition follows from the definition of induced faces and the
first assertion. To prove the sufficiency, assume that $F$ is a
maximal face and that $\dim\cR(\G(F))=d-1$ for some $\G\in\T$. We have to show that the face $F$ is induced. Without any loss of generality we may assume that $\G$ is the identity, i.e., that $\dim\cR(F)=d-1$. As $F$ is maximal, \eqref{eq:FandF'} implies that $F=F_{\cR(F)}$  and so $F$ is an induced face of $\cS_1$.
 \epf

To illustrate Proposition \ref{pp:maxfacedim}, we mention three examples.  The face $\D_b$ in Example \ref{ex:FinF'} is not maximal, see \cite[p. 147]{hk14}. For any face $F$ containing 
$\D_b$ we have $\G(F)\supseteq\G(\D_b)=\D_b$ for all $\G\in\T$ 
which implies that $\cR(\G(F))=\cH$. 
Let $\ket{z_i}$, $i=1,\ldots,10$ be as in that example. For $k\in\{1,\ldots,10\}$ let $F_k$ denote the maximal face of $\D_b$ not containing the vertex $\proj{z_k}$. Since any nine vectors 
$\ket{z_i}$ span $\cH$, we have $\cR(F_k)=\cH$ for each $k$. If $F'$ is a maximal face of $F_k$ then $\cR(F')\subset\cH$ is a hyperplane, but $F'$ is not a maximal face of $\cS_1$.

The next two examples refer to the maximal faces of two qubits which we shall construct in Proposition \ref{pp:twoqubit} below. For $F=F_{V_1}$ we have $\dim\cR(\G(F))=d-1$ for all $\G\in\T$. For $F=F_{V_2}$ we have $\dim\cR(F)=d-1$ while
$\dim\cR(\G_1(F))=d$.

There is a proper non-maximal face $F$ of $\cS_1$ such that
$\cR(F)=\cH$ and $F$ is not a polytope. For example, we can take  $F$ to be the convex hull of the product states $\proj{x,x^*}$ with $\norm{x}=1$ and $d_1>2$. Evidently $F\in\cF$. One can verify that $\cR(F)=\cH$ and $\dim \cR(\G(F))<d-1$ for all 
$\G\in\T$ and $n=2$. It follows from Proposition \ref{pp:maxfacedim} that $F$ is not maximal.

Let us also comment on the recent paper \cite{hk2013} where the
authors have constructed in $2\ox4$ a face $F\in\cF$ which is the
convex hull of ten points $\proj{z_1}$, $\proj{z(\a_i)}$,
$i=2,\ldots,10$. Moreover, one has $\dim \cR(F)\le 7$. As each
hyperplane of $\cH$ contains infinitely many product vectors,
Observation 3 and Proposition \ref{pp:maxfacedim} imply that $F$ is not maximal.


\textbf{Open problem 3}. \quad Can a maximal face of $\cS_1$ be a polytope?

The next lemma follows easily from \cite[Theorem 4]{as10}.
 \bl
 \label{le:prod}
Let $\ket{a}=\ket{a_1,\ldots,a_n}$ and
$\ket{b}=\ket{b_1,\ldots,b_n}$ be non-parallel product vectors
with $\|a_i\|=\|b_i\|=1$ for each $i$, and let $F\in\cF$ be the
face generated by $\proj{a}$ and $\proj{b}$. If the vector
$\ket{a}+\ket{b}$ is entangled, then $F$ is the straight line
segment joining $\proj{a}$ and $\proj{b}$. Otherwise, there is a
unique index $i$ such that $\ket{a_i}$ and $\ket{b_i}$ are
non-parallel, and $F=\{\proj{x}\}$ where
$\ket{x}=\ket{a_1,\ldots,a_{i-1},x_i,a_{i+1},\ldots,a_n}$ and
$\ket{x_i}$ runs through all unit vectors in $\lin\{a_i,b_i\}$.
 \el

The above lemma implies that $\dim F>0$ for any maximal face
$F\in\cF$. This leads to the following problem.

\textbf{Open problem 4}. \quad For a fixed quantum system, find the
minimum and the maximum of $\dim F$ over all maximal faces $F\in\cF$. In particular, are there any faces of dimension $d^2-2$?

In the bipartite case it is known that there is no face of dimension
$d^2-2$ (see \cite{gL07}).

It follows from Proposition \ref{pp:maxfacedim} that the faces constructed in Propositions \ref{pp:SymOper} and
\ref{pp:RealVersion} are not maximal except in the two-qubit case (see Proposition \ref{pp:twoqubit}). Hence, in the case
$d_1=\cdots=d_n$ the maximum mentioned in the above problem is bigger than the dimension \eqref{eq:dimF}.

\subsection{Maximal faces associated to hyperplanes} \label{sec:hyperplane}

The following theorem provides a very rich family of maximal faces
of $\cS_1$, namely the faces $F_V$ where $V\su\cH$ is any
hyperplane. Recall that if $V\su\cH$ is a subspace of codimension
$>1$ then the associated face $F_V$ is not maximal (see Proposition \ref{pp:maxfacedim}). 

 \bt
 \label{thm:hyperplane}
For a subspace $V\sue\cH$, the associated face $F_V$ is maximal if and only if $V$ is a hyperplane.
 \et
  \bpf
In view of Proposition \ref{pp:maxfacedim}, we need only to prove
the sufficiency part. For any product vector
$\ket{\a}=\ket{a_1,a_2,\ldots,a_n}$ and distinct indexes
$i_1,\ldots,i_s\in\{1,\ldots,n\}$ and any  vectors
$\ket{x_{i_k}}\in\cH_{i_k}$ we denote by
$\a(x_{i_1},\ldots,x_{i_s})$ the product vector obtained from
$\ket{\a}$ by replacing each $\ket{a_{i_k}}$ with the corresponding $\ket{x_{i_k}}$.

Let $F\in\cF$ be such that $F_V\subset F$. Since this inclusion is strict, we have $P_F\not\sue V$. We claim that for any product vector $\ket{\a}=\ket{a_1,a_2,\ldots,a_n}\in P_F\setminus V$ with
$\norm{a_1}=\cdots=\norm{a_n}=1$, any distinct indexes
$i_1,\ldots,i_s\in\{1,\ldots,n\}$ and any unit vectors
$\ket{x_{i_k}}\in\cH_{i_k}$ we have
$\a(x_{i_1},\ldots,x_{i_s})\in P_F$. The proof will be by
induction on $s=1,\ldots,n$.

The case $s=1$ is easy. We may assume that $\ket{x_{i_1}}$ and  $\ket{a_{i_1}}$ are linearly independent. Since $V$ is a hyperplane there exists a unit vector
$\ket{v_{i_1}}\in\lin\{ \ket{a_{i_1}},\ket{x_{i_1}} \}$ such that
$\a(v_{i_1})\in V$. Since both $\ket{\a}$ and
$\a(v_{i_1})$ belong to $P_F$, Lemma \ref{le:prod} implies that
also $\a(x_{i_1})\in P_F$.

Now assume that the claim is true for some $s<n$. We set
$\ket{\b}=\a(x_{i_{s+1}})$. By using the
case $s=1$ we have $\ket{\b}\in P_F$. If $\ket{\b}\notin V$
then the induction hypothesis implies that
$\b(x_{i_1},\ldots,x_{i_s})\in P_F$, i.e.,
$\a(x_{i_1},\ldots,x_{i_{s+1}})\in P_F$. Next assume that
$\ket{\b}\in V$. Then $\ket{a_{i_{s+1}}}$ and $\ket{x_{i_{s+1}}}$ are linearly independent and we denote by $\ket{y_{i_{s+1}}}$ a unit vector parallel to $\ket{a_{i_{s+1}}}+\ket{x_{i_{s+1}}}$. Note that $\ket{\g}:=\a(y_{i_{s+1}})\in P_F\setminus V$. By the
induction hypothesis we have
$\a(x_{i_1},\ldots,x_{i_s})\in P_F$ as well as
$\g(x_{i_1},\ldots,x_{i_s})\in P_F$. It follows from
Lemma \ref{le:prod} that
$\a(x_{i_1},\ldots,x_{i_s},x_{i_{s+1}})\in P_F$. Hence, the
claim holds also for $s+1$.

We conclude that the claim is valid for all $s=1,\ldots,n$.
When $s=n$, the claim implies that $F=\cS_1$ and so $F_V$ must be  a maximal face of $\cS_1$.
 \epf

Recall that all induced faces of $\cS_1$ are exposed. Whether this is true for arbitrary faces is apparently not known.

\textbf{Open problem 5}. \quad Is every face of $\cS_1$ exposed?

 \bcr \label{cr:HyperplaneSection}
If $V\subset\cH$ is a hyperplane then $P_V$ spans $V$.
 \ecr
 \bpf
Let $V'\sue V$ be the subspace spanned by $P_V$. Then
$F_{V'}=F_V$, and the theorem implies that $F_{V'}$ is a maximal
face. It follows from the theorem that $V'$ must be a hyperplane,
and so $V'=V$.
 \epf

We make two remarks related to this corollary. First, we remark that there exist subspaces of $\cH$ of codimension $2$ which are not spanned by product vectors. An example is the subspace spanned by the entangled vector 
$(\ket{01}+\ket{10})\otimes\ket{0,\ldots,0}$ and the $d-3$ basic product vectors $\ket{i_1,i_2,i_3,\ldots,i_n}$ subject to the 
condition that $(i_1,i_2)\notin\{(0,0),(0,1),(1,0)\}$ if 
$i_3=0,\ldots,i_n=0$.

Second, one can view the decomposable vectors in the fermionic space $\we^N\bC^M$ as the counterparts of product vectors. Recently, it has been shown that there exists a subspace of $\we^N\bC^M$ of codimension $3$ which is not spanned by decomposable $N$-vectors \cite[Proposition 14]{ccz13}. In the same reference, it has been proved that when $N=2$ any subspace of codimension at most two is spanned by decomposable $N$-vectors. Whether this is true for $N>2$ is still unknown.

An interesting problem is to compute the dimension of the maximal faces $F_V$, where $V$ is a hyperplane. Two ILO-equivalent hyperplanes give rise to maximal faces of the same dimension. Therefore it suffices to consider only the representatives of the ILO-equivalence classes of hyperplanes. If $n=2$ there are only finitely many equivalence classes. Apart from the bipartite systems, the number of equivalence classes of hyperplanes is finite only in finitely many cases, all of them 3-partite. As a first step, we shall compute the dimension of the maximal face associated to the hyperplane orthogonal to a product vector.

 \bpp  \label{pp:concrete}
If $\ket{\a}=\ket{a_1,a_2,\ldots,a_n}\in\cH$ is a product vector,
then the dimension of the maximal face $F_V$ associated to the
hyperplane $V=\ket{\a}^\perp$ is given by the formula
 \bea \label{eq:DimFormula}
\dim F_V = d^2 - 1 -\prod_{i=1}^n (2d_i-1).
 \eea
 \epp
 \bpf
Denote by $H_i$ the space of Hermitian operators on $\cH_i$. Let $H'_i=\{\r_i\in H_i:\cR(\r)\perp\ket{a_i}\}$ and let $H''_i=(H'_i)^\perp\su H_i$. Note that $\dim H_i=d_i^2$, $\dim H'_i=(d_i-1)^2$ and $\dim H''_i=2d_i-1$. A product vector  belongs to $V$ if and only if it belongs to one of the subspaces $\cH_1\ox\cdots\ox\cH_{i-1}\ox\ket{a_i}^\perp\ox\cH_{i+1}\ox\cdots\ox\cH_n$. It follows that $\dim F_V=\dim L -1$ where $L=\sum_{i=1}^n H_1\ox\cdots\ox H_{i-1}\ox H'_i\ox H_{i+1}\ox H_n$. This sum is not a direct sum but it may be also written as a direct sum, namely $L=\oplus_{i=1}^n H''_1\ox\cdots\ox H''_{i-1}\ox H'_i\ox H_{i+1}\ox\cdots\ox H_n$. Hence
 \bea
\dim L &=& \sum_{i=1}^n \frac{2d_1-1}{d_1^2}\cdots
\frac{2d_{i-1}-1}{d_{i-1}^2}
\left(1-\frac{2d_i-1}{d_i^2}\right) d^2 \notag \\
&=& d^2 \left( 1-\prod_{i=1}^n \frac{2d_i-1}{d_i^2} \right)
 \eea
and the formula \eqref{eq:DimFormula} follows.
 \epf

In the bipartite case $(n=2)$, with $d_1\le d_2$, there are exactly
$d_1$ ILO-equivalence classes of hyperplanes $V\su\cH$. Their
representatives are $V_j=\ket{\ps_j}^\perp$, $j=1,\ldots,d_1$, where
$\ket{\psi_j}\in\cH$ is any vector of Schmidt rank $j$, e.g.,
$\ket{\psi_j}=\sum_{i=1}^j\ket{i,i}$. The dimension of $F_{V_1}$ has been computed in the above proposition. We leave aside the problem of computing in general the dimension of $F_{V_j}$ for $j>2$. Here we compute $\dim F_{V_2}$.

 \bpp \label{pp:Rank-2}
Let $n=2$ and $V=\ket{\ps}^\perp$, where $\ket{\ps}\in\cH$ is a
vector of Schmidt rank 2. Then $\dim F_V=d(d-2)$.
 \epp
 \bpf
We may assume that $\ket{\psi}=\ket{0,1}-\ket{1,0}$. We denote by $S_i$ $(i=1,2)$ the subspace of $\cH_i$ spanned by the first two basis vectors $\ket{0}$ and $\ket{1}$. We also set $W_i=S_i^\perp\su\cH_i$. For convenience, we shall identify $S_1$ and $S_2$
and denote this space by $S$. Then any product vector in $V$ can be written as $(t_1\ket{x}+\ket{z})\ox(t_2\ket{x}+\ket{y})$,
where $\ket{x}\in S$, $\ket{z}\in W_1$, $\ket{y}\in W_2$ and
$t_i$ are real. Let $L\sue H$ be the subspace spanned by all product states
$(t_1\ket{x}+\ket{z})(t_1\bra{x}+\bra{z})\ox(t_2\ket{x}+\ket{y})(t_2\bra{x}+\bra{y})$.
As $t_i$ are real parameters, $L$ is spanned by all Hermitian
matrices of the following nine types:

1) $\proj{x}\ox\proj{x}$,

2) $(\ketbra{x}{z}+\ketbra{z}{x})\ox\proj{x}$,

3) $\proj{x}\ox(\ketbra{x}{y}+\ketbra{y}{x})$,

4) $(\ketbra{x}{z}+\ketbra{z}{x})\ox(\ketbra{x}{y}+\ketbra{y}{x})$,

5) $\proj{z}\ox\proj{y}$,

6) $\proj{z}\ox\proj{x}$,

7) $\proj{x}\ox\proj{y}$,

8) $(\ketbra{x}{z}+\ketbra{z}{x})\ox\proj{y}$,

9) $\proj{z}\ox(\ketbra{x}{y}+\ketbra{y}{x})$.

Thus, $L=\oplus_{i=1}^9 L_i$ where $L_i$ is the subspace spanned by the matrices of type $i)$. We have $\dim L_1=9$,
$\dim L_5=(d_1-2)^2(d_2-2)^2$, $\dim L_6=4(d_1-2)^2$,
$\dim L_7=4(d_2-2)^2$, $\dim L_8=4(d_1-2)(d_2-2)^2$ and
$\dim L_9=4(d_1-2)^2(d_2-2)$. It remains to show that
$\dim L_2=12(d_1-2)$, $\dim L_3=12(d_2-2)$ and
$\dim L_4=14(d_1-2)(d_2-2)$. These three proofs use the same
arguments and we shall prove only that
$\dim L_4=14(d_1-2)(d_2-2)$.

$H$ is a real subspace of the complex vector space $M$ of all $d\times d$ matrices. Let $T$ be the real subspace of $M$ consisting of all upper triangular matrices with real diagonal elements. The map $H\to T$ which sends a Hermitian matrix to its upper triangular part is an isomorphism of real vector spaces. Denote by $L'_4$ the image of $L_4$ by this isomorphism. Let us write $\ket{x}=\sum_{j\in\{0,1\}} \xi_j\ket{j}$,
$\ket{z}=\sum_{1<l<d_1} \zeta_l\ket{l}$ and
$\ket{y}=\sum_{1<k<d_2} \eta_k\ket{k}$. Then
 \bea
 (\ketbra{x}{z}+\ketbra{z}{x})\ox(\ketbra{x}{y}+\ketbra{y}{x})
 &=&
 \sum(
 \xi_j\zeta_l^*\xi_r\eta_k^* \ketbra{j}{l}\ox\ketbra{r}{k}+
 \xi_j\zeta_l^*\xi_r^*\eta_k \ketbra{j}{l}\ox\ketbra{k}{r}
 \notag\\
 && \quad
+\xi_j^*\zeta_l\xi_r\eta_k^* \ketbra{l}{j}\ox\ketbra{r}{k}+
 \xi_j^*\zeta_l\xi_r^*\eta_k \ketbra{l}{j}\ox\ketbra{k}{r}),
 \eea
where the summation is over all $j,r\in\{0,1\}$,
$l\in\{2,3,\ldots,d_1-1\}$, and $k\in\{2,3,\ldots,d_2-1\}$. Thus,
$L'_4$ is the real subspace of $T$ spanned by all matrices
 \bea
 \sum_{j,r=0}^1 \sum_{l=2}^{d_1-1} \sum_{k=2}^{d_2-1} (
 \xi_j\zeta_l^*\xi_r\eta_k^* \ketbra{j}{l}\ox\ketbra{r}{k}+
 \xi_j\zeta_l^*\xi_r^*\eta_k \ketbra{j}{l}\ox\ketbra{k}{r}).
 \eea
This sum has $8(d_1-2)(d_2-2)$ terms but only $7(d_1-2)(d_2-2)$ different monomials occur. By Lemma \ref{le:Monomials} these monomials are linearly independent over complex numbers. Since each of them takes real as well as imaginary values, it follows from Corollary \ref{cr:MonDim} that
$\dim L_4=\dim L'_4=14(d_1-2)(d_2-2)$.
This completes the proof.
 \epf

In the remainder of this section we shall consider in more details two low-dimensional systems.

\subsection{Two-qubit system} \label{sec:2q}

We consider here the two-qubit system only. Note that in this case we have $\cS=\cP$ as well as $\cS_1=\cP_1$. The faces of $\cP$ were classified in \cite{hk04}. Their classification relies
on the previous work \cite{bk02} where the faces of the cone of
positive maps on the algebra of complex $2\times2$ matrices
were classified. It was observed in \cite{Kye} that some maximal 
faces have dimension 8. We present here a new and simple method to classify the maximal faces of $\cS_1$ up to the $\diamond$-action. This agrees with the classification in \cite{hk04}. Let us start with the description, in the case of two qubits, of the face $F$ constructed in Proposition \ref{pp:SymOper}.

 \bex \label{ex:Face2q} {\rm
Let $V\subset\cH$ be the hyperplane consisting of all symmetric
tensors in $\cH$. The set $P_V$ consists of all product vectors
$\ket{x,x}$ with $\|x\|=1$. The associated face $F_V$ is the convex hull of all $\proj{z}$ with $\ket{z}\in P_V$. Explicitly,
this face consists of all positive semidefinite matrices
 \bea \label{eq:SymFace}
\left[ \begin{array}{cccc}
a   & x   & x   & z \\
x^* & b   & b   & y \\
x^* & b   & b   & y \\
z^* & y^* & y^* & c
\end{array} \right],\quad a+2b+c=1.
 \eea
Obviously, $\dim F_V=8$ and $F_V$ is not a polytope. By Theorem
\ref{thm:hyperplane} it is a maximal face of $\cS_1$. By Proposition
\ref{pp:SymOper}, the extreme points of $F_V$ are the matrices
 \bea \label{eq:ExtP}
\left[ \begin{array}{cccc}
|\xi|^4 & |\xi|^2\xi\eta^* & |\xi|^2\xi\eta^* & (\xi\eta^*)^2 \\
|\xi|^2\xi^*\eta&|\xi\eta|^2&|\xi\eta|^2&|\eta|^2\xi\eta^*\\
|\xi|^2\xi^*\eta&|\xi\eta|^2&|\xi\eta|^2&|\eta|^2\xi\eta^*\\
(\xi^*\eta)^2 & |\eta|^2\xi^*\eta & |\eta|^2\xi^*\eta & |\eta|^4
\end{array} \right],\quad |\xi|^2+|\eta|^2=1.
 \eea
The face $F=F_V^{\G_1}$ is still maximal. It is easy to verify that almost all $\r\in F$ have full rank, and so we have $\cR(F)=\cH$.
 \qed }\eex

We can now determine all maximal faces of $\cS_1$.
 \bpp
 \label{pp:twoqubit}
In the case of two qubits, there are only three $\GL$-equivalence
classes of maximal faces of $\cS_1$. Their representatives are:
$F_{V_1}$, $F_{V_2}$, $F_{V_2}^{\G_1}$, where
$V_1=\ket{0}\otimes\cH_2+\cH_1\otimes\ket{0}$ and $V_2$ is the
space of symmetric tensors in $\cH$.
 \epp
 \bpf
It follows from Theorem \ref{thm:hyperplane} that the three
representatives are indeed maximal faces. Let $F$ be any maximal
face. It follows from Proposition \ref{pp:maxfacedim} that both $\cR(F)$ and $\cR(F^{\G_1})$ have dimension at least $d-1=3$. Assume that both dimensions are 4. Then there exist 
$\r_1,\r_2\in F$ such that
both $\r_1$ and $\r_2^{\G_1}$ have rank 4. Hence the state
$(\r_1+\r_2)/2\in F$ is a full state on $\pars$. However, we have
shown that when $d\le6$ there are no such states on $\pars$. This
means that at least one of $\cR(F)$ and $\cR(F^{\G_1})$ must have
dimension 3. As $\G_1$ interchanges the faces $F_{V_2}$ and
$F_{V_2}^{\G_1}$,  and preserves $F_{V_1}$, we may assume that
$V:=\cR(F)$ has dimension 3. Hence, \eqref{eq:F_AV} implies that
$F=F_V$. For nonzero $\ket{a}$ we have
$V\cap(\ket{a}\otimes\cH_2)\ne0$. If $\ket{a}\otimes\cH_2\sue V$ for
some nonzero $\ket{a}$, then it is easy to show that $V$ is
$\GL$-equivalent to $V_1$. It follows that $F$ is $\GL$-equivalent
to $F_{V_1}$. If there is no such $\ket{a}$, then there is an
$A\in\GL$ such that $AV$ is the space of symmetric tensors. Hence,
$F$ must be $\GL$-equivalent to $F_{V_2}$ and our claim is proved.
 \epf

Note that $\dim F_{V_1}=6$ and $\dim F_{V_2}=8$ and so we have the following corollary.
 \bcr \label{cr:MaxFaces2x2}
In $2\otimes2$, the minimum and the maximum of $\dim F$ over all maximal faces $F$ is $6$ and $8$, respectively.
 \ecr

\subsection{Qubit-qutrit system} \label{sec:2x3}

In this section we describe the maximal faces of $\cS_1$ in the
$2\otimes3$ system. As a byproduct, we obtain that the maximum
dimension of the proper faces is 24.

 \bpp
 \label{pp:2x3system}
In the qubit-qutrit system, there are only three
$\GL$-equivalence classes of maximal faces of $\cS_1$. Their
representatives are: $F_{V_1}$, $F_{V_2}$, $F_{V_2}^{\G_1}$, where $V_1=\ket{0,0}^\perp$ and
$V_2=(\ket{0,2}-\ket{1,0})^\perp$.
 \epp
 \bpf
It follows from Theorem \ref{thm:hyperplane} that the three
representatives are indeed maximal faces. Let $F$ be any maximal
face. It follows from Proposition \ref{pp:maxfacedim} that both $V:=\cR(F)$ and $\cR(F^{\G_1})$ have dimension at least $d-1=5$. The case where both dimensions are 6 can be ruled out by the same argument as in the two-qubit case. Thus at least one of $\cR(F)$ and $\cR(F^{\G_1})$ must have dimension 5. As $\G_1$ fixes 
$F_{V_1}$ and interchanges $F_{V_2}$ and $F_{V_2}^{\G_1}$, we may assume that $\dim V=5$. As $F$ is maximal, \eqref{eq:FandF'} implies that $F=F_V$. Note that $\dim V\cap(\ket{a}\otimes\cH_2)\ge2$ for any
$\ket{a}\ne0$.

Assume first that there is a nonzero vector $\ket{a}$ such
that $\ket{a}\otimes\cH_2\subset V$. Then there exists
$A\in\GL$ such that $AV=V_1$, and so $F$ is $\GL$-equivalent to $F_{V_1}$. From now on we assume that
$\ket{a}\otimes\cH_2\not\subset V$ for all $\ket{a}\ne0$, and
so $\dim V\cap(\ket{a}\otimes\cH_2)=2$. One can easily show that
there is a basis $\{\ket{a_i}:i=0,1\}$ of $\cH_1$ and a basis
$\{\ket{b_j}:j=0,1,2\}$ of $\cH_2$ such that $V$ is spanned by
the product vectors $\ket{a_0,b_0}$, $\ket{a_0,b_1}$,
$\ket{a_1,b_1}$, $\ket{a_1,b_2}$, and $\ket{a,b}$ where
$\ket{a}=\ket{a_0}+\ket{a_1}$ and
$\ket{b}=\ket{b_0}+\ket{b_1}+\ket{b_2}$. Consequently, there
exists $A\in\GL$ such that $AV=V_2$ and we obtain that $F$ is
$\GL$-equivalent to $F_{V_2}$.

Since $F_{V_1}$ is $\GL$-equivalent $F_{V_1}^{\G_1}$, it is
clear that $F_{V_1}$ is $\GL$-equivalent to neither $F_{V_2}$
nor $F_{V_2}^{\G_1}$. One can verify that
$\cH_{F^{\G_1}_{V_2}}=\cH$. Hence, $F_{V_2}$ is not
$\GL$-equivalent to $F_{V_2}$.
 \epf

 \bcr \label{cr:MaxFaces2x3}
In $2\otimes3$, the minimum and the maximum of $\dim F$ over all maximal faces $F$ is $20$ and $24$, respectively.
 \ecr
 \bpf
It suffices to observe that $\dim F_{V_1}=20$ by Proposition 
\ref{pp:concrete} and $\dim F_{V_2}=24$ by Proposition \ref{pp:Rank-2}.
 \epf

From these results we have the following observation.
 \bpp
 \label{pp:minimum}
Each maximal face of $\cS_1$ (of any multipartite quantum system)
has dimension at least $6$. This lower bound is saturated in the
case of two qubits.
 \epp
 \bpf
Suppose there is a maximal face $F$ of dimension $<6$. It
follows from Proposition \ref{pp:maxfacedim} that $\dim F\ge d-2$, and so $d\le 7$. As $n>1$ and each $d_i>1$, we have $n=2$ and
we may assume that $d_1=2$ and $d_2=2,3$. As we have computed
the dimensions of the maximal faces in these two cases, we
deduce that this is impossible. Thus the first assertion holds.
For the second assertion see subsection \ref{sec:2q}.
 \epf

\section{Some EW for two qutrits} \label{sec:3x3}

In this section we study the remarkable family of faces $\D_b$ of $\cS_1$ in the two-qutrit quantum system, depending on the parameter $b>0$, $b\ne1$. This family was constructed by Ha and Kye in \cite{hk14}. We have mentioned this family in Example \ref{ex:FinF'} and referred to it already several times. In particular, we have shown in that example that the above faces
$\D_b$ are not induced. We point out that the very recent paper \cite[p. 14]{hk13e}, by the same authors, contains another family of faces in $3\ox3$ having similar properties and depending on two real parameters. We shall consider only the one-parameter family in this section. We shall include in our treatement the 
three limiting cases $b=0,1,\infty$.

\subsection{A cyclic inequality} \label{sec:Cyclic}

The following inequality plays a crucial role in the
proof of the main result of this section, Proposition 
\ref{pp:witness}.

 \bl \label{le:PomNej}
Let $a,b>0$, $c\ge0$, $a\ge2(b+c)-3\sqrt{bc}$ and
$(ab-c^2)(ac-b^2)<0$. Then the homogeneous cyclic inequality
 \bea \label{eq:CikNej}
\frac{x}{ax+by+cz}+\frac{y}{ay+bz+cx}+\frac{z}{az+bx+cy}
\le\frac{3}{a+b+c}
 \eea
holds when $x,y,z\ge0$ and $x+y+z>0$. The equality sign holds
only if (i) $x=y=z$ or at the points (ii) $x=0,bz^2=cy^2$; (iii)
$y=0,bx^2=cz^2$; (iv) $z=0,by^2=cx^2$ when $c>0$ and
$a=2(b+c)-3\sqrt{bc}$.
 \el
 \bpf
Note that $b\ne c$ because $(ab-c^2)(ac-b^2)<0$. It is easy to
check that
 \bea \label{eq:ProstaN}
2(b+c)-3\sqrt{bc}>\frac{b^2+c^2}{b+c},
 \eea
and so $a(b+c)>b^2+c^2$. As $(b^2+c^2)/(b+c)>\sqrt{bc}$, we also
have $a^2>bc$. Let $f(x,y,z)$ denote the function on the left hand
side of the inequality \eqref{eq:CikNej}. We shall first examine the
critical ponts of $f$ in the interior of the first orthant, i.e.,
when $x,y,z>0$. The partial derivatives $f_x,f_y,f_z$ of $f$ vanish
at a cirtical point $(x,y,z)$. By taking the linear combinations
$bf_y-cf_x$ and $bf_z-cf_y$, we obtain the equations
$f_1f_2=0,~f_3f_4=0$ where
 \bea
f_1 &=& z(bx-cy)+cx^2-by^2,\\
f_2 &=& (a^2-bc)(bx+cy)+(2abc-b^3-c^3)z,\\
f_3 &=& x(by-cz)+cy^2-bz^2, \\
f_4 &=& (a^2-bc)(by+cz)+(2abc-b^3-c^3)x.
 \eea
If $f_1=0$ it is easy to verify that $bx-cy\ne0$, and so
$z=(by^2-cx^2)/(bx-cy)$. If $f_1=f_3=0$ then we also have $x=y=z$
and $f(x,y,z)=3/(a+b+c)$. If $f_1=f_4=0$ we obtain that
$y(ab-c^2)^2=x(ac-b^2)^2$ and
 \bea \label{eq:nej-1}
\frac{3}{a+b+c}-f(x,y,z)=\frac{-(a(b+c)-(b^2+c^2))^3}
{(a+b+c)(ab-c^2)(ac-b^2)(a^2+b^2+c^2-ab-ac-bc)}>0.
 \eea

If $f_2=0$ then we can solve this equation for $x$. We must have
$f_3=0$ or $f_4=0$. In both cases we obtain again that
\eqref{eq:nej-1} holds.

It remains to consider the boundary of the first orthant. The
inequality holds strictly if two of the varables $x,y,z$ vanish. For
instance, if $y=z=0$ then $a\ge(b^2+c^2)/(b+c)>(b+c)/2$ because
$b\ne c$.

Assume that only $z=0$. The inequality holds strictly if $c=0$.
Otherwise we have
$f(x,y,0)=g(t)=(c+2at+bt^2)/((a+bt)(c+at))$ where $t=y/x$. The
function $g(t)$ has a maximum at $t=\sqrt{c/b}$ and at that point
we have
 \bea
\frac{3}{a+b+c}-g \left( \frac{\sqrt{c}}{\sqrt{b}} \right) = 
\frac{a-2(b+c)+3\sqrt{bc}}{(a+b+c)(a+\sqrt{bc})}\ge0.
 \eea
Thus, the equality sign holds in \eqref{eq:CikNej} only if $c>0$,
$a=2(b+c)-3\sqrt{bc}$ and $by^2=cx^2$.

The other two cases can be treated similarly.
 \epf

\subsection{Entanglement witnesses} \label{sec:EW}

Let us recall the definition of (optimal) entanglement witnesses 
in the bipartite case.

 \bd \label{def:EW}
A Hermitian operator $W\in H$ is an
{\em entanglement witness (EW)} if it satisfies the following two
conditions

(i) $W$ has at least one negative eigenvalue;

(ii) $\tr(W\r)\ge0$ for all $\r\in\cS_1$.

Thus, if $\r\in\cD_1$ and $\tr(W\r)<0$ then $\r$ is an entangled
state. An entanglement witness $W$ is an {\em optimal entanglement witness (OEW)} if the set of
entangled states detected by $W$ is maximal, i.e.,

(iii) there is no entanglement witness $W'$ such that
$\{\rho\in\cD_1:\tr(W\r)<0\}$ is a proper subset of
$\{\rho\in\cD_1:\tr(W'\r)<0\}$.
 \ed

We say that an EW, say $W$, has the {\em spanning property} if
$\cH$ is spanned by the product vectors $\ket{z}$ such that
$\tr(W\proj{z})=0$. It is a well known fact (see
\cite[Corollary 2]{lkc00}) that an EW satisfying the spanning property is optimal. This is a sufficient but not necessary
condition for optimality. On the other hand, a necessary and
sufficient condition for optimality says that an EW $W$ is optimal if and only if $W-P$ is not an EW for any nonzero  positive semidefinite matrix $P$ (see \cite[Theorem 1]{lkc00}). We will use both criteria in the sequel. If $W$ is an EW, then the condition (ii) implies that $F_W$ is a proper exposed face of $\cS_1$. (For the notaion $F_W$ and $X_W$ see 
subsection \ref{subsec:TheFace}.) 
If $W$ is an OEW, then the face $F_W$ is nonempty.

From now until the end of this section we consider only the 
quantum system $3\ox3$. Let us introduce a 1-parameter family $\{W_b\}$, $b\in[0,\infty]$, of normalized (i.e., with trace 1) Hermitian operators. It is given by the formula
 \bea \label{eq:MatW}
W_b &=& \frac{1}{4}I_9
-\frac{(1+b)^2}{12(1-b+b^2)} \sum_{i=1}^6 \proj{z_i}
-\frac{3(1-3b+b^2)}{16(1-b+b^2)} \sum_{i=7}^{10} \proj{z_i},
 \eea
where the $\ket{z_i}$ are the normalized product vectors from
\cite[p. 17]{hk14}. For $i=1,3,5$ these product vectors are given by
 \bea
\ket{z_1} &=& ( \ket{0}+ \sqrt{b} \ket{1} ) \otimes
( \sqrt{b} \ket{0}+ \ket{1} ) /(1+b), \\
\ket{z_3} &=& ( \ket{1}+ \sqrt{b} \ket{2} ) \otimes
( \sqrt{b} \ket{1}+ \ket{2} ) /(1+b), \\
\ket{z_5} &=& ( \ket{2}+ \sqrt{b} \ket{0} ) \otimes
( \sqrt{b} \ket{2}+ \ket{0} ) /(1+b).
 \eea
For $i=2,4,6$ they are given by the same formulas except
that one should replace $\sqrt{b}$ with $-\sqrt{b}$. The
remaining four are
 \bea
\ket{z_7} &=& (\ket{0}+\ket{1}+\ket{2})\otimes
(\ket{0}+\ket{1}+\ket{2})/3, \\
\ket{z_8} &=& (\ket{0}+\ket{1}-\ket{2})\otimes
(\ket{0}+\ket{1}-\ket{2})/3, \\
\ket{z_9} &=& (\ket{0}-\ket{1}+\ket{2})\otimes
(\ket{0}-\ket{1}+\ket{2})/3, \\
\ket{z_{10}} &=& (-\ket{0}+\ket{1}+\ket{2})\otimes
(-\ket{0}+\ket{1}+\ket{2})/3.
 \eea
By definition, $W_\infty=\lim_{b\to\infty} W_b$. In order to simplify notation, we have supressed the fact that the $\ket{z_i}$ for $i\le6$ depend on the parameter $b$. We shall write $X_b$ for the hyperplane $X_{W_b}$. 

We denote by $\D_b$ the convex hull of the ten states $\proj{z_i}$. When $b\ne0,\infty$, one can easily verify that the ten states $\proj{z_i}$ are linearly independent and so $\D_b$ is a 9-dimensional simplex. In the two exceptional cases, $b=0,\infty$, we have $\ket{z_i}=\ket{z_{i+1}}$ for $i=1,3,5$ and $\D_b$ becomes a 6-dimensional simplex.

For $b\ne0,1,\infty$, it was shown in \cite{hk14} that $\D_b$
is a face of $\cS_1$. In the next proposition we show that all $W_b$ are EW (they are OEW for $b\ne1$), and we shall write $F_b$ for the face $F_{W_b}$. We will show that $F_b=\D_b$ for $b\ne1$. 
In particular we obtain another proof of the result mentioned 
above.

 \bpp  \label{pp:witness}
Let $W_b,X_b,F_b$, and the $\ket{z_i}$ be as defined above.

(i) If $b\ne0,1,\infty$ then $W_b$ is an OEW, $F_b=\D_b$, and
$W_b$ has the spanning property.

(ii) For $b=1$, $W_1$ is a non-optimal EW and $F_1$ is the convex
hull of all states $\proj{x,x}$ with real $\ket{x}\in\cH_1$
and $\|x\|=1$.

(iii) If $b=0,\infty$, then $W_b$ is an OEW, $F_b=\D_b$, but $W_b$ lacks the spanning property.
 \epp
 \bpf
(i) The characteristic polynomial of $W_b$ has the factorization
 \begin{equation} \label{eq:KarPol}
2^{-8}(t+b)(2t-3+5b-3b^2)^2 \left( 4t^2-4(1+b^2)t-1+2b+b^2
+2b^3-b^4 \right)^3.
 \end{equation}
Thus $-b$ is a negative eigenvalue of $W_b$ and so the condition (i) of Def. \ref{def:EW} is satisfied. In order to verify the condition (ii) of the same definition, it suffices to prove that the inequality $\tr(W_b\r)\ge0$ holds for all pure product states
$\r=\proj{x,y}$, $\ket{x}=\sum_i x_i\ket{i}\in\cH_1$ and
$\ket{y}=\sum_i y_i\ket{i}\in\cH_2$.
A computation gives
 \bea
 \label{eq:real} \notag
 6(1-b+b^2)\tr(W_b\r) &=&
 (1-b)^2\sum_i |x_iy_i|^2 + b^2\sum_i |x_iy_{i+1}|^2
 +\sum_i |x_iy_{i-1}|^2 \\
 && -2(1-b+b^2)\sum_i \Re(x_i^* x_{i+1}) \Re(y_i^* y_{i+1}).
 \eea
For convenience, the subscript $i$ runs through integers modulo 3 and we shall use this convention in the rest of this proof.
Thus, we have to show that the bihermitian form in the
variables $x_i$ and $y_i$ on the right hand side is positive semidefinite. We can view
this form as a hermitian form in the $y_i$. Then our task reduces
to showing that the matrix of this hermitian form is positive
semidefinite, i.e.,
 \bea \label{eq:MatX}
X(b) &:=& \left[ \begin{array}{ccc}p_0&0&0\\0&p_1&0\\0&0&p_2
\end{array} \right]
-\frac{1-b+b^2}{2}\left( \proj{x}+ \proj{x^*} \right) \ge0,
 \eea
where $p_i=(2-3b+2b^2)|x_i^2|+|x_{i+1}^2|+b^2|x_{i+2}^2|>0$. This matrix inequality can be re-written as
 \bea \label{eq:MatNejP}
P^{-1/2}\proj{x}P^{-1/2}+P^{-1/2}\proj{x^*}P^{-1/2}\le
\frac{2}{1-b+b^2} I_3,
 \eea
where $P=\diag(p_0,p_1,p_2)$. Since
$\bra{x^*}P^{-1}\ket{x^*}=\bra{x}P^{-1}\ket{x}=
\sum_i |x_i|^2/p_i$, it suffices to prove that
 \bea \label{eq:NejP}
\frac{|x_0^2|}{p_0}+\frac{|x_1^2|}{p_1}+\frac{|x_2^2|}{p_2}
\le\frac{1}{1-b+b^2}.
 \eea
It is straightforward to verify that this inequality follows from Lemma \ref{le:PomNej}. (The $a,b,c,x,y,z$ of that lemma should be set to $2-3b+2b^2,1,b^2,|x_0^2|,|x_1^2|,|x_2^2|$, respectively.)
Thus we have shown that $W_b$ is an EW.
As $\tr(W_b\proj{z_i})=0$ for $i=1,\ldots,10$, and the
$\ket{z_i}$ span $\cH$, we conclude that $W_b$ has the spanning property and so it is an OEW.

In order to prove that $F_b=\D_b$, it suffices to show that
$\tr(W_b\proj{x,y})=0$ implies that $\ket{x,y}\propto\ket{z_i}$
for some $1\le i\le10$. Thus, let us assume that
$\tr(W_b\proj{x,y})=0$. Then the equality sign must hold in the
inequality \eqref{eq:NejP}. By Lemma \ref{le:PomNej}, there are
four possibilities.

First, the $x_i$ have the same modulus, say 1.
We can assume that $x_0=1$, $x_1=e^{i\a}$, $x_2=e^{i\b}$ with
$\a,\b\in\bR$. By plugging these values into $X(b)$, we find that
 $$
\det X(b)=6(1-b+b^2)^3 (3-\cos 2\a-\cos 2\b-\cos2(\a-\b)).
 $$
As $X(b)$ must be singular, we deduce that
$\cos 2\a=\cos 2\b=\cos2(\a-\b)=1$. Thus we obtain that
$x_1=\pm1$ and $x_2=\pm1$. In each of these four subcases
$X(b)$ has rank 2 and so there is (up to a scalar factor) a
unique vector $\ket{y}$ such that $X(b)\ket{y}=0$. Therefore
$\ket{x,y}\propto\ket{z_i}$ for some $i=7,\ldots,10$.

Second, $x_0=0$ and $|x_2|^2=b|x_1|^2$. We may assume that
$x_1=1$ and so $|x_2|=\sqrt{b}$. Since $X(b)$ must be a
singular matrix, we obtain that $x_2=\pm\sqrt{b}$. This
leads to the solutions $\ket{z_3}$ and $\ket{z_4}$.

The remaining two cases are similar to the last one.
Thus, we have shown that $F_b=\D_b$.

(ii) Let $\ket{x,y}\in\cH$ be an arbitrary unit product vector.
Then one can verify that
 \bea \label{eq:W'-ravan} \notag
12\bra{x,y} W_1 \ket{x,y}
 &=&
|x_0y_1-x_1y_0|^2+|x_0y_2-x_2y_0|^2+|x_1y_2-x_2y_1|^2 \\
&& +|x_0y_1^*-x_1y_0^*|^2+|x_0y_2^*-x_2y_0^*|^2+
|x_1y_2^*-x_2y_1^*|^2,
 \eea
where $x_i$ and $y_i$ are the components of $\ket{x}$ and $\ket{y}$,
respectively. It follows that $W_1$ is an EW and that $F_1$
is indeed the convex hull of the set of pure product states
$\proj{x,x}$ with real $\ket{x}\in\cH_1$ and $\|x\|=1$. However, $W_1$ is not an OEW. Indeed, if $\ket{e_{ij}}=\ket{ij}-\ket{ji}$, then one can easily verify that
 \bea
W&:=&2W_1-\frac{1}{6}\sum_{0\le i<j<3} \proj{e_{ij}} \notag \\
&=& \frac{1}{6} \left( I_9-\sum_{i,j=0}^2 \ketbra{ii}{jj} \right)
 \eea
is an EW.

(iii) We give the proof only when $b=0$. The case $b=\infty$ 
can be treated similarly. (Note that $W_b=W_{1/b}$ for all 
$b\in[0,\infty]$ and that $\ket{z_i(1/b)}$ is obtained from 
$\ket{z_i(b)}$ by switching $\cH_1$ and $\cH_2$.)
By setting $b=0$ in Eq. \eqref{eq:KarPol} we infer that $W_0$ has a negative eigenvalue, namely $(1-\sqrt{2})/2$. By using the continuity of the map sending $b\to W_b$, we deduce that $W_0$ is an EW. To prove the optimality, we need to show that $W_0-P$ is not an EW for any nonzero $P\ge0$ \cite{lkc00}. Since the product states $\proj{z_i}\in X_0$, we may assume that $P\ket{z_i}=0$. Thus, $\cR(P)\sue\lin\{\ket{\a},\ket{\b}\}$ where $\ket{\a}=\ket{00}-\ket{11}$ and $\ket{\b}=\ket{11}-\ket{22}$. We have $P=\proj{p\a+q\b}+\proj{r\a+s\b}$, where $p,r\ge0$ and $q,s$ are  some complex numbers. By a straightforward computation, for any $b>0$ we have
$\bra{z_1}W_0-P\ket{z_1}=-b(|q|^2+|s|^2-b/6)/(1+b)^2$
and $\bra{z_3}W_0-P\ket{z_3}=-b(p^2+r^2-b/6)/(1+b)^2$. Hence, at
least one of these expressions is negative for small $b>0$.
We conclude that $W_0$ is an OEW.

Next we show that $W_0$ violates the spanning property. Since $b=0$, we have $\ket{z_i}=\ket{z_{i+1}}$ for $i=1,3,5$. Hence, it suffices to show that if $\ket{z}=\ket{x,y}$ and $\bra{z}W_0\ket{z}=0$ then $\ket{z}\propto\ket{z_i}$ for some $i$. If a component of $\ket{x}$ or $\ket{y}$ vanishes, we claim that $\ket{z}\propto\ket{z_i}$ for some $i\in\{1,3,5\}$. Say $x_0=0$, then \eqref{eq:real} with $b=0$ can be rewritten as $\abs{x_1y_1-x_2y_2}^2+\abs{x_1^*y_1-x_2^*y_2}^2+2\abs{x_1y_0}^2+2\abs{x_2y_1}^2=0$. It follows that $\ket{z}\propto\ket{z_3}$ or $\ket{z}\propto\ket{z_5}$. This proves our claim. Assume now that no component of $\ket{x}$ or $\ket{y}$ vanishes. Let $\ket{x'}\in\cH_1$ be the vector having components $|x_i|$, and define $\ket{y'}\in\cH_2$ similarly. We also set $\ket{z'}=\ket{x',y'}$. By setting $b=0$ in \eqref{eq:real} and by using the fact that $W_0$ is an EW, we obtain that 
\bea  \label{eq:nonzero}
0=6\bra{z}W_0\ket{z}\ge
\sum_i |x_iy_i|^2 + \sum_i |x_iy_{i-1}|^2 
-2\sum_i \abs{x_i x_{i+1}y_i y_{i+1}} 
=6\bra{z'}W_0\ket{z'}\ge0.
\eea
It follows that $\abs{x_i x_{i+1}y_i y_{i+1}}=\Re(x_i^* x_{i+1}) \Re(y_i^* y_{i+1})$ for $i=0,1,2$. Since no component of $\ket{x}$ or $\ket{y}$ vanishes, both $x_i^* x_{i+1}$ and $y_i^* y_{i+1}$ are real. Without any loss of generality we may assume that all 
$x_i$ and $y_i$ are real. Moreover, we may assume that $x_0>0$, $y_0>0$, $x_1y_1>0$ and $x_2y_2>0$. Define the matrix $X(0)$ by setting $b=0$ in \eqref{eq:MatX}. It was shown in the proof of part (i) that $X(b)\ge0$ for $b>0$. By continuity, we also have $X(0)\ge0$. An easy computation gives $\bra{y}X(0)\ket{y}=6\bra{z}W_0\ket{z}=0$. As $\ket{y}\ne0$, it follows that $\det X(0)=0$. Equivalently, when $b=0$ then the equality holds in \eqref{eq:NejP}. By applying Lemma \ref{le:PomNej} and assuming that 
$\|x\|=\|y\|=1$, we obtain that $x_0^2=x_1^2=x_2^2=1/3$. It follows easily that $\ket{z}\propto\ket{z_i}$ for some $i\in\{7,8,9,10\}$. This completes the proof.
 \epf
 
Let us make a few additional remarks about the faces $F_b$ 
in the above proposition. We claim that the faces $F_0,F_1$ (and $F_\infty$) are not induced. If we set $n=2$ and $d_1=d_2=3$ in Proposition \ref{pp:RealVersion}, then $F'=F_1$, and so $F_1$ is not induced. The proof for $F_0$ is similar to the proof of part (iii) of the mentioned proposition. It uses the facts that the subspace $\cR(F_0)$ is spanned by 7 linearly independent real product vectors and that this subspace contains infinitely many product vectors. By Proposition \ref{pp:maxfacedim} the faces $F_0,F_1,F_\infty$ are not maximal. For other faces $F_b$ see the 
example below that proposition.

 \bcr
When $d_1=d_2=3$ then the set of normalized OEW is not closed.
 \ecr
 \bpf
It suffices to note that $W_1$ is the limit of the sequence
$W_{1+1/m}$ as $m\to\infty$.
 \epf



\section*{Competing interests}

The authors declare that no competing interests exist.

\section*{Authors contributions}

L.C. and D.D. drafted the manuscript and approved the study 
for publication.


\section*{Acknowledgments}

We thank S.-H. Kye for his comments and for providing the references \cite{hk14,hk13e,Kye}. We also thank an anonymous 
referee for his comments.

\section*{Funding statement}

L.C. was partially supported by the Fundamental Research Funds for the Central Universities (Grant Nos. 30426401 and 30458601). 
D.Z.D. was supported in part by NSERC (Canada).


\begin{thebibliography}{99}

\bibitem{cd13JMP} Lin Chen and D. \v{Z}. {\Dbar}okovi{\'c},
2013 Dimensions, lengths, and separability in finite-dimensional
quantum systems,
\jmp {\bf54}, 022201.


\bibitem{ja13} P. D. Jarvis, 2013 The mixed two qutrit system: 
local unitary invariants, entanglement monotones, and the SLOCC
group $\SL(3,\bC)$,  quant-ph/1312.7413v1.


\bibitem{as10} E. Alfsen and F. Shultz, 2010 Unique decompositions, faces, and automorphisms of separable states, J. Math. Phys. 51, 052201.


\bibitem{hk14} K.-C. Ha and S.-H. Kye, 2014 Separable states
with unique decompositions, 
Comm. Math. Phys. (2014), 328, 131--153.

\bibitem{hk2014} K.-C. Ha and S.-H. Kye, 2014 Construction of exposed indecomposable positive linear maps between matrix algebras, 
arXiv:1410.5545v1 [math.OA] 21 Oct 2014


\bibitem{it06} L. M. Ioannou and B. C. Travaglione, 2006 Quantum
separability and entanglement detection via entanglement-witness
search and global optimization, \pra {\bf73}, 052314.


\bibitem{lan12} J. M. Landsberg, 2012 Tensors: Geometry and
Applications, Amer. Math. Soc., Providence, R. I.


\bibitem{Harris:1992}
J. Harris, 1992 Algebraic Geometry, A First Course,
Springer, New York.

\bibitem{cd13CMP} Lin Chen and D. \v{Z}. {\Dbar}okovi{\'c},
2013 Properties and construction of extreme bipartite states having positive partial transpose,
Comm. Math. Phys. {\bf 323}, 241--284.

\bibitem{horodecki97} P. Horodecki, 1997 Separability criterion and inseparable mixed states with positive partial transpose, Phys. Lett. A {\bf232}, 333--339.

\bibitem{werner89} R. F. Werner, 1989 \pra {\bf 40}, 4277.

\bibitem{hh99} M. Horodecki and P. Horodecki, 1999 \pra {\bf59}, 4206.

\bibitem{cd12PRA} Lin Chen and D. \v{Z}. {\Dbar}okovi{\'c}, 2012 Qubit-qudit states with positive partial transpose,
\pra {\bf86}, 062332.

\bibitem{dc00} W. D\"{u}r and J. I. Cirac, 2000 Classification of multiqubit mixed states: Separability and distillability properties,
\pra, {\bf 61}, 042314.

\bibitem{kck00} B. Kraus, J. I. Cirac, S. Karnas, and M. Lewenstein, 2000 \pra {\bf 61}, 062302.

\bibitem{bh67} B. Huppert, 1967 Endliche Gruppen I, Springer,
New York.

\bibitem{hk2013} K.-C. Ha and S.-H. Kye, 2013 Geometry for separable
states and construction of entangled states with positive partial
transposes, \pra {\bf88}, 024302.

\bibitem{gL07} O. G\"{u}hne and N. L\"{u}tkenhaus, 2007,
J. Phys.: Conf. Ser. {\bf67}, 012004.

\bibitem{ccz13} Jianxin Chen, Lin Chen, and Bei Zeng, 2013 Unextendible product basis for fermionic systems, 
arXiv:1312.4218.

\bibitem{hk04} K.-C. Ha and S.-H. Kye, 2004 Construction of
entangled states with positive partial transposes based on indecomposable positive linear maps,
Phys. Lett. A {\bf325}, 315--323.

\bibitem{bk02} E.-S. Byeon and S.-H. Kye, 2002 Facial structures
for positive linear maps in two-dimensional matrix algebra,
Positivity {\bf 6}, 369--380.

\bibitem{Kye} S.-H. Kye, 2013 Faces for two-qubit separable states and the convex hulls of trigonometric moment curves, Prob. Math. Stat. {\bf 33}, 385--400.

\bibitem{hk13e}  K.-C. Ha and S.-H. Kye, 2013 Exposedness of Choi 
type entanglement witnesses and applications to lengths of 
separable states, 
Open Systems \& Information Dynamics, {\bf 20} 
(doi: 10.1142/S1230161213500121).
quant-ph/1211.5675v3 (2013).

\bibitem{lkc00} M. Lewenstein, B. Kraus, J. I. Cirac, and P.
Horodecki, 2000 Optimization of entanglement witnesses, 
\pra {\bf62}, 052310.



%
%
%
%
%
%
%
%
%
%
%
%
%
%
%
%


\end{thebibliography}
\end{document}